\documentclass[pdflatex,sn-nature]{sn-jnl}

\usepackage{graphicx}
\usepackage{multirow}
\usepackage{amsmath,amssymb,amsfonts}
\usepackage{amsthm}
\usepackage{mathrsfs}
\usepackage[title]{appendix}
\usepackage{xcolor}
\usepackage{textcomp}
\usepackage{manyfoot}

\usepackage{booktabs}
\usepackage{tabularx}
\usepackage{array}
\usepackage{ragged2e}
\newcolumntype{L}{>{\RaggedRight\arraybackslash}X}
\usepackage[most]{tcolorbox}
\usepackage{tikz}
\usetikzlibrary{decorations.pathreplacing,arrows.meta}
\newcommand{\cbox}{$\square$\ }

\begin{document}

\title[Data readiness for agentic AI in the Earth sciences]{From FAIR to AWARE: Data Readiness for Agentic AI in the Earth Sciences}

\author*[1]{\fnm{Ivan} \sur{Kuznetsov}}\email{ivan.kuznetsov@awi.de}
\equalcont{These authors contributed equally to this work.}

\author*[1]{\fnm{Dmitrii} \sur{Pantiukhin}}\email{dmitrii.pantiukhin@awi.de}
\equalcont{These authors contributed equally to this work.}

\author[1]{\fnm{Sonja} \sur{H\"anzelmann}}

\author[2]{\fnm{Fabian} \sur{Wachsmann}}

\author[3]{\fnm{Jonas} \sur{S{\o}lvsteen}}

\author[3]{\fnm{Daniel} \sur{Wiesmann}}

\author[1,4]{\fnm{Thomas} \sur{Jung}}

\author[1]{\fnm{Nikolay} \sur{Koldunov}}

\affil[1]{\orgname{Alfred Wegener Institute, Helmholtz Centre for Polar and Marine Research}, \orgaddress{\city{Bremerhaven}, \country{Germany}}}

\affil[2]{\orgname{Deutsches Klimarechenzentrum (DKRZ)}, \orgaddress{\city{Hamburg}, \country{Germany}}}

\affil[3]{\orgname{Development Seed}, \orgaddress{\city{Lisbon}, \country{Portugal}}}
\affil[4]{Institute of Environmental Physics, University of Bremen, Bremen, Germany}

\abstract{Artificial intelligence in the Earth sciences is beginning to act: prototype agents already search archives, run simulations and draft analyses. Scientific data were curated for researchers who bring expert judgement; an autonomous agent cannot be assumed to supply it. Data can satisfy the FAIR principles and still mislead a machine acting alone. We propose that Earth-science data become AWARE (Actionable, Well-described, Auditable, Recipe-equipped, Epistemically labelled), an agent-specific requirements layer that extends FAIR, within a three-part agenda: AWARE data, bounded tools, layered validation. The ingredients exist across community infrastructure. We outline a staged, measurable path to adoption, intended to reduce silent error.}

\maketitle

\section{Introduction}

In a recent controlled evaluation by Zhou and colleagues \cite{egusphere2026}, an autonomous artificial-intelligence agent was asked to analyse the water balance of large river basins. It loaded the supplied data, wrote the analysis code, and produced a polished figure with values within a few per cent of the reference, close enough to pass visual inspection. The figure was wrong: the agent had silently dropped the land-fraction weighting. No exception was raised, and nothing in the output revealed the omission. The necessary information was not missing: the agent received the same descriptions of the data that the tool-based condition used, and its script even located the model's cell-area variable. Nothing, however, required the agent to act on that knowledge. When the same underlying large language models (LLMs) were instead restricted to a catalogue of validated, self-describing tools, the share of benchmark tasks solved correctly on the first attempt rose from about 5\% to 82\% (seven hydrology tasks, graded bit-for-bit against expert-verified references) \cite{egusphere2026}. The gain is consistent with broader evidence that LLMs equipped with validated tools and retrieval over curated knowledge outperform the same models operating alone \cite{schick2023,lewis2020,yang2024}. The size of the effect, however, comes from a single study.

This Perspective examines the source of this discrepancy and the changes to data infrastructure needed to close it. The failure described above occurred in a controlled evaluation; as agentic AI enters routine scientific practice, the same silent failure mode will recur across many analyses. The evaluation also shows that validated tools prevent such errors; we argue that the knowledge those tools encode must ultimately reside with the data, in versioned, machine-actionable form.

The problem is urgent because adoption is running ahead of governance. A growing literature debates how generative artificial intelligence should be governed in science, from authorship and ethics \cite{natportfolio_ai,cope2023,icmje} to assessment procedures \cite{ipccai2026}. Meanwhile, researchers already use LLMs throughout the workflow: drafting methods, searching the literature and the archives, processing data, and writing analysis code. Agentic AI is moving inside the scientific loop: LLM agents now plan and execute sequences of actions on a user's behalf through natural-language interfaces.

Three concurrent advances drive this shift. The first is the maturation of machine-learning models of the Earth system itself: data-driven weather prediction has moved from a curiosity to systems that match or exceed selected operational benchmarks at a small fraction of the inference cost \cite{lam2023,bi2023,pathak2022,price2024,alet2025}. The approach has since spread to foundation models of the atmosphere and ocean, hybrid and climate-scale emulation, and kilometre-scale downscaling \cite{bodnar2025,nguyen2023,schmude2024,kochkov2024,wattmeyer2025,chen2024fuxi,allen2025,mardani2025}. The second is the emergence of LLMs tuned to geoscientific vocabulary and literature \cite{thulke2024,bi2024,deng2024,lin2024,webersinke2021}. Together they mean that fast, callable models of the Earth system now exist for an agent to invoke the way it invokes any other tool. 

The third advance is \emph{agentic orchestration}. Whereas the first two advances improved existing components, orchestration changes how they are used: instead of answering questions, multi-agent systems plan, call tools, execute code in sandboxed environments, inspect their own figures, and correct their own errors across multi-step workflows. Such systems, most of them still research prototypes, already automate the analysis loop over archives at petabyte scale \cite{guo2025,kim2025} and ground themselves in curated knowledge graphs to reproduce published results \cite{jaber2025}. Others discover and analyse data across multidisciplinary archives \cite{pantiukhin2025}, deliver localized climate information through an openly released service prototype \cite{koldunov2024,kuznetsov2025}, and unify Earth-observation analysis through an ecosystem of more than one hundred tools \cite{feng2026}.

Large institutional programmes support this development. The European Union's Destination Earth initiative launched its initial operational system, including the first two digital twins of the Earth system, in June 2024; a full digital replica is targeted for 2030 \cite{ec_destine,hoffmann2026}. The Helmholtz Association funds Earth-system foundation-model pilots \cite{helmholtz_hfmi}. In the United States, the National Science Foundation funds an institute for trustworthy artificial intelligence in weather, climate and coastal oceanography \cite{mcgovern2022}, and the National Oceanic and Atmospheric Administration (NOAA) operates a centre for artificial intelligence \cite{noaa_ncai}; the European Space Agency (ESA) runs comparable programmes for Earth observation \cite{esa_philab}. In one Scopus bibliometric sample, machine-learning publications on climate grew by more than two orders of magnitude between 2008 and 2022 \cite{ajibade2023}. These programmes invest chiefly in models, digital twins and platforms; the readiness of the data their agents will act on has attracted no comparable investment.

Systematic evaluations, however, show that these systems remain unreliable. On ScienceAgentBench, which comprises 102 data-analysis tasks extracted from 44 peer-reviewed publications in four disciplines, the best evaluated agent solved 32.4\% of tasks given three attempts each \cite{chen2024sab}. On DSBench, 466 data-analysis tasks drawn from real data-science competitions, the best agent solved 34.1\% \cite{jing2024}. More revealing than the failure rate is the nature of the failures. On TerraBench, which poses 403 tasks over heterogeneous Earth-system data (gridded fields, satellite imagery, geospatial context and simulator output) and grades tool selection and numerical accuracy separately, agents did far better at choosing the right tool than at producing numerically acceptable answers: more than 84\% of numerical results fell outside tolerance, and the grounding of tool arguments (region, period, parameters) was a principal failure source \cite{nguyen2026}. That benchmark, at least, locates the failures in data handling rather than scientific reasoning; the broader benchmarks establish capability limits without resolving their cause. Much of the knowledge that would prevent such errors is not published; it is held within individual research groups: which datasets can be validly compared, which regridding method conserves the global integral of a variable, and which simulation versions carry known defects. Model scaling alone cannot guarantee access to knowledge that is unpublished, local, or specific to a dataset version.

These observations motivate the central argument of this Perspective: the orchestration layer is maturing faster than the data layer, and agentic systems amplify the weaknesses of the resources on which they operate. This assessment is informed by our own development and operation of agentic systems over Earth-science archives \cite{koldunov2024,kuznetsov2025,pantiukhin2025,pantiukhin2026}. We argue that agent readiness rests on three pillars: (1) data that carry their own expert context, formalized below as the \textbf{AWARE} requirements; (2) an infrastructure of composable tools and standard machine interfaces that agents can operate safely; and (3) layered validation, since full autonomy is not a realistic near-term goal: pre-execution code checks, runtime assertions and adversarial review to catch errors before release, and machine-readable provenance to keep released results re-executable.

The unconstrained agent of our opening example had none of these safeguards; the tool-restricted condition supplied the second pillar alone, and accuracy rose accordingly. Agent readiness requires all three pillars, and each depends on skilled human labour (curation, tool development, review) that must be credited and funded, a governance question addressed below. The remainder of this Perspective first examines why Earth-science data resist autonomous use, then develops the three pillars, shows that their components already exist within community infrastructure, and outlines a staged path to adoption.

\section{Why Earth-science data are not ready for autonomous agents}

Earth-science data can satisfy the FAIR principles (findable, accessible, interoperable, reusable) \cite{wilkinson2016} and still be unusable by autonomous agents. Machine actionability is an explicit goal of FAIR, but the principles are deliberately general: they do not prescribe domain-specific validity constraints or executable handling instructions, and they assume a reader who supplies tacit knowledge wherever documentation is ambiguous. A fully autonomous agent is the limiting case of the ``distant reuser'' \cite{borgman2025}: it has no way to consult the data creator. Some of what such consultation conveys is tacit and cannot be written down at all \cite{polanyi1966,collins2010}; this Perspective concerns the part that can be made explicit. Five structural properties of Earth-science data widen the gap between FAIR compliance and agent usability, and each motivates one of the AWARE requirements introduced in the next section, named in parentheses below.

\textbf{The decoding burden} (answered by \emph{Recipe-equipped})\textbf{.} Earth-science data are stored in several distinct encoding families, each externalizing decoding knowledge differently. The Network Common Data Form (NetCDF) family, the backbone of model archives \cite{rew1990}, is largely self-describing: its modern variants reside inside a hierarchical container (HDF5) whose header declares the compression filters and chunking in use. The header does not, however, supply the decoder itself: filters are plugins resolved from a registry outside the file, so an agent in an environment lacking the required plugin can fail on valid files; such failures are uncommon but difficult to diagnose. The GRIB format of operational meteorology is a stronger example: it is interpretable only through versioned code tables maintained by the World Meteorological Organization \emph{outside} the file. Cloud-native formats such as Zarr expose metadata that agents can parse natively, but exist in multiple specification versions with extensible compression registries \cite{zarr_spec}; the same pattern holds elsewhere: seismic waveforms require external station metadata before instrument responses can be removed, and ocean profile quality-control flags are defined in reference tables maintained separately from the profiles they describe.

\textbf{Semantics are voluntary} (answered by \emph{Well-described})\textbf{.} Community metadata conventions supply the semantic layer that gives arrays their scientific meaning. In climate science this role falls to the Climate and Forecast (CF) conventions, which define variable identity, physical units, calendars and grid descriptions and have been formalized in an encoding-independent data model \cite{hassell2017}. But enforcement of such conventions is uneven across projects and repositories, and real archives contain a long tail of partially compliant data. A human reader routes around a missing or ambiguous attribute; an agent that assumes compliance mishandles it silently. Model grids provide a canonical example. The same physical variable may be delivered on a regular latitude--longitude grid by one model, on a curvilinear ocean grid with two-dimensional coordinate arrays by a second, and on an unstructured mesh by a third; if the metadata describing the grid is incomplete, all three still load as ordinary arrays, but any regional selection, spatial average or model intercomparison computed from them is quietly wrong. Agents will encounter such incomplete metadata in practice; robustness to partial compliance is therefore a core requirement.

\textbf{Grids are silent-correctness traps} (answered by \emph{Well-described, Recipe-equipped})\textbf{.} Regular, rotated-pole, curvilinear and unstructured-mesh grids all decode into ordinary-looking arrays. Earth-observation imagery adds projection, datum and bounding-box semantics that agents must interpret correctly before any spatial operation is valid. Regridding a variable whose area integral must be preserved, using a method that does not conserve it, corrupts the budget without raising any exception; taking an unweighted average over a regular latitude--longitude grid overweights the poles; ignoring cell-area information, external land--sea masks or missing-value sentinels produces plausible numbers that are quantitatively wrong. The companion files that prevent these errors are themselves a weak link. The CF conventions can declare the dependency (through \texttt{cell\_measures} and \texttt{external\_variables}), but archives populate these attributes unevenly (in CMIP6, cell-area fields such as \texttt{areacella} are separate datasets, absent for some model--experiment combinations), and the convention specifies no resolution mechanism. In our audited sessions (28 incidents from four agentic systems operated by our group, catalogued in the Supplementary Information), an agent located the correct cell-area dataset and then did not use it, and a reanalysis land mask was borrowed from a climate model, contaminating the comparison along complex coastlines. Neither failure raised an error. These are the highest-severity failure modes for autonomous pipelines precisely because they are invisible in the output. Humans make some of these errors too; the difference lies in scale: an agent can repeat one mistake across a thousand analyses without collegial review.

\textbf{Scale and structure} (answered by \emph{Actionable, Auditable})\textbf{.} The sixth phase of the Coupled Model Intercomparison Project (CMIP6), the most recent with a fully populated archive (the seventh is under way), was designed around 312 experiments across 23 endorsed sub-projects, with an archive volume estimated at roughly twenty petabytes \cite{eyring2016,balaji2018,petrie2021}. The archive is published through the Earth System Grid Federation (ESGF), a deliberately decentralised system of dozens of independently administered nodes in which replication, versioning, authentication and node outages are constant operational concerns \cite{cinquini2014,petrie2021}. Dataset identity is governed by a formal reference syntax and controlled vocabularies \cite{taylor_drs}, and even the set of available variables and output frequencies is a formal, versioned object \cite{juckes2020}. At this scale ``download then process'' is no longer viable: agents must default to lazy, chunked, server-side access through remote-access protocols, cloud-optimized stores and virtual aggregations \cite{cimadevilla2025}. 

Federation failures are particularly hazardous in combination with agent autonomy: when access to part of an ensemble fails mid-analysis, an agent may silently substitute other data. For exchangeable realizations of the same model this can be statistically harmless; substituting a different model configuration is not, and either way the change must be declared. In one audited session, an agent that lost access to a high-resolution configuration substituted an alternative and logged the change (incident D3, Supplementary Information); the log was correct, but the ensemble's composition, and hence its meaning, had shifted. Without a declared substitution policy, such drift surfaces only in a post-hoc audit of the logs.

\textbf{Epistemic class is invisible} (answered by \emph{Epistemically labelled})\textbf{.} A free-running CMIP6 projection, an ERA5 reanalysis field \cite{hersbach2020}, an interpolated Argo profile and a satellite retrieval all decode into identical-looking multidimensional arrays, yet they are different kinds of scientific object. The day-to-day weather of a free-running simulation does not correspond to the weather that occurred on those dates, and a reanalysis blends a model with observations, so it is neither ground truth nor independent of models and carries assimilation artefacts and its own uncertainty. Nothing in the files records these distinctions, so no file checker can warn an agent that applies the wrong kind of data to a task. In our audited runs, a pipeline subtracted a reanalysis baseline from future model output and interpreted the difference as climate change. That difference, however, also contains the model's systematic bias with respect to the reanalysis; the standard approach is to subtract the same model's historical simulation. Both datasets were individually correct, convention-compliant and FAIR; the subtraction itself was the invalid operation, and it was caught only in review. Valid-use constraints must therefore travel with the data; in the AWARE checklist they appear twice, as the epistemic label that governs comparisons and as a separate machine-readable valid-use record.

\section{From FAIR to AWARE data}

The first pillar concerns the data itself. Three notions of data readiness must be distinguished: \textbf{FAIR}, \textbf{AI/ML-ready} and \textbf{agent-ready}; the third is the least developed. \textbf{FAIR} \cite{wilkinson2016} establishes general findability, accessibility, interoperability and reuse, with machine actionability an explicit goal; it deliberately leaves domain-specific validity open. \textbf{AI/ML-ready} adds what a training pipeline needs: the Croissant format makes datasets directly loadable into ML frameworks \cite{akhtar2024}, FAIR-for-AI extends the principles to models and benchmarks \cite{huerta2023}, and FAIR Digital Objects reframe data as typed, operable units \cite{schultes2019,desmedt2020}. 

Agent-ready adds what neither of the first two provides: \emph{the expert context that a human silently supplies at the moment of reuse}, externalized into machine-actionable form so far as it can be stated, bound to the data by a persistent identifier and resolvable without human intervention. Such context is rarely recorded: in a survey of 4,041 data papers, collection and source information was commonly documented, while generalization limits appeared in 8\% of papers and maintenance policies in under 1\% \cite{giner2025}.

The theoretical basis is the ``data creators' advantage'' \cite{pasquetto2019}: effective reuse depends on tacit knowledge that humans obtain by collaborating with data creators. A fully autonomous agent cannot rely on timely collaboration, so the context must ship with the data, as ``Datasheets for Datasets'' argued for human consumers \cite{gebru2021}, but as structured assertions rather than prose. Where the assertions run out, the right agent behaviour is to stop and escalate to the data steward; the contract should make such escalation the exception by answering the recurring questions once for all agents. The cost of leaving this context unpublished is illustrated by one archive, where supporting natural-language discovery required rewriting 1,768 terse variable, experiment and model descriptions into self-contained plain-language entries \cite{pantiukhin2026}. Much of that work reconstructed, and partly had to infer, context that data producers are best placed to publish once. This enrichment can be partly automated: LLMs can draft the entries for experts to review.

We refer to data that close this gap as \textbf{AWARE}:
\begin{itemize}\setlength{\itemsep}{2pt}
  \item[] \textbf{A}ctionable: every reuse-critical element resolves to a machine endpoint, never to a human-only web page; licences and access are machine-negotiable;
  \item[] \textbf{W}ell-described: semantics are checked rather than merely recommended, with standardized variable names, machine-parseable physical units, calendars, grid and coordinate-system topology, masks and cell areas, plus a structured register of known artefacts;
  \item[] \textbf{A}uditable: versioned, checksummed identity with machine-readable lineage following the W3C PROV standard and a citation bundle linking the exact version to the literature that used it;
  \item[] \textbf{R}ecipe-equipped: a deterministic, executable recipe for loading and correctly handling the data ships with the data itself; conceptually, a recipe is a \emph{skill} the dataset supplies to any agent: a packaged, validated competency that agents invoke rather than reimplement;
  \item[] \textbf{E}pistemically labelled: the dataset declares what it \emph{is} (free-running model output, assimilated reanalysis, satellite retrieval, in-situ observation) and what it may validly be compared against; CF's free-text \texttt{source} attribute anticipates the class label, but agents need it as a controlled, checkable vocabulary together with the comparison rules.
\end{itemize}
The acronym summarizes the requirement: data that carry their own operating instructions, history and limits are a necessary part of any architecture for autonomous use, though not sufficient on their own. Together the five requirements address the structural problems of the preceding section, several of which are addressed jointly, as the labels there indicate. AWARE does not replace FAIR: FAIR sets general principles, and AWARE adds the agent-specific requirements a domain must specify before a reuser who cannot ask questions can act on the data.

The requirements are not specific to gridded data; gridded data are in fact the more tractable case. Model output and reanalyses have regular structure, and their conventions, however unevenly enforced, at least exist. Observational data are different. They come from thousands of instruments, sensors and campaigns, and in many communities each provider packages the results as it sees fit before uploading them to a database. There is often no convention an agent could even assume, so every structural choice must be guessed from whatever metadata happen to be present. A human expert resolves such heterogeneity by inspection; for an agent, each guess is a silent-failure risk, and at archive scale repeated guessing also wastes computation and energy. Observational data therefore stand to gain the most from AWARE: a shipped load recipe and declared conventions replace guesswork exactly where no shared convention exists.

CMIP6 illustrates both the extent and the limits of current metadata enrichment. The infrastructure was designed to connect each published dataset to systematic documentation of the model and experiment that produced it \cite{pascoe2020}, to a dedicated citation service for evolving data \cite{stockhause2017}, and to an errata service that tracks known issues. But the documentation link embedded in a dataset resolves to a page written for human readers, not to a machine-actionable document. Context held on a separate service is, moreover, accessible only while that service remains online; when it is unavailable or discontinued, the agent loses the context with it. For an agent to act on the knowledge that a specific simulation carries, say, a known sea-ice defect over a particular period, that knowledge must resolve as a structured assertion (affected component, severity, temporal bounds, corrective action) rather than as narrative text.

The hardest ingredient of AWARE data, expert judgement about \emph{fitness for purpose}, is already produced at scale, but it is published in forms an agent cannot act on. The European Space Agency's Climate Change Initiative, for example, publishes systematic product validation and intercomparison reports for its essential-climate-variable datasets \cite{hollmann2013}, and comparable evaluation functions operate at other major providers. These assessments contain the tacit-context layer an autonomous agent needs, but they are published as prose, figures and reports rather than as version-bound, machine-validated assertions. Recast as structured assertions bound to the dataset identifier, they would let agents select data by quality and use case rather than by name alone; the translation requires scientific judgement and governance. The situation is general: an archive that invests in quality assessment, errata tracking or user guidance already possesses the raw material of AWARE data and lacks only a machine-readable rendering of the knowledge it curates.

Fitness for purpose also has a statistical dimension that current metadata rarely states: monthly means cannot support daily-exceedance metrics, and small single-realization ensembles may be underpowered to resolve scenario differences in regional extremes, yet nothing warns an agent which class of question a dataset can answer; the dependence runs through the signal-to-noise ratio: the same single realization that answers a question about the mean state may be useless for regional extremes. A register of such statements could become arbitrarily detailed; a workable contract will start from the few classes of question that recur (means, trends, extremes) rather than aim for completeness.

The same holds for the record of who has used a dataset and how. Standards for dataset--literature links, provenance description and research-object packaging together form the vocabulary of a \emph{citation bundle} \cite{burton2017,moreau2013,soiland2024}, yet today an agent cannot generally recover a complete, version- and variable-level path from a dataset identifier to the papers that used it, the methods they applied, and the caveats they reported. Agentic systems that need this grounding currently build it themselves, wiring corpora of thousands of publications into every analysis \cite{jaber2025,pantiukhin2026}. A community knowledge graph binding dataset identity to provenance and linked literature at the granularity of individual variables and versions would therefore be a productive investment. Grounding of this kind also guards against fabricated or misidentified datasets: in our audited runs, verifying citations against retrieved sources caught fabricated DOIs (Supplementary Information).

No existing catalogue plays the role of the authoritative contract. The SpatioTemporal Asset Catalog serves Earth-observation scenes, and faceted model-archive catalogues serve gridded output; the contract must add typed, validatable records on top of either. Two gaps recur. First, discovery metadata rarely states, as a consistent typed field, what a dataset is good for and what it is not; humans carry that knowledge, and in practice agent interfaces end up reconstructing it by hand. Second, licences illustrate the Actionable gap. Terms are named in metadata, but acceptance remains a per-dataset click on a website, which halts a fully autonomous retrieval until a human intervenes. Designing machine-readable terms with a delegable, auditable acceptance path is tractable, provided acceptance remains attributable to a responsible person or institution. Box~1 condenses the requirements; Supplementary Note~1 sketches a minimal machine-readable contract.

\begin{tcolorbox}[breakable,colback=gray!5,colframe=black!70,title=\textbf{Box 1 $|$ The AWARE checklist for agent-ready Earth-science data (implementation of the first pillar)}]
A dataset is AWARE when a machine-actionable contract, bound to a versioned persistent identifier, provides:
\begin{itemize}\setlength{\itemsep}{3pt}
  \item[\cbox] \textbf{Resolvable identity \& version} \textit{(Auditable)}: PID resolves to a specific immutable version with checksum; replica and node state discoverable; ``latest'' is never implicit.
  \item[\cbox] \textbf{Self-contained load recipe} \textit{(Recipe-equipped)}: format, required decoding components, chunking layout and access protocol declared machine-readably; a reference invocation with its expected output lets any agent self-test the recipe.
  \item[\cbox] \textbf{Checked semantics} \textit{(Well-described)}: every variable carries a resolvable standard name, UDUNITS-parseable units, and explicit calendar/time reference; existing convention checkers (CF-checker, PrePARE) already automate this at publication time.
  \item[\cbox] \textbf{Grid \& operation contract} \textit{(Well-described)}: grid topology declared (regular, rotated, curvilinear or unstructured mesh; map projection and datum for Earth-observation data); cell areas and land--sea masks present, or resolvably referenced, when the declared operation requires them; the invariant each operation must preserve stated per variable and declared purpose (a budget calculation must conserve the area integral; a quick-look map need not), with the safe choice as default and faster paths as logged overrides.
  \item[\cbox] \textbf{Epistemic class label} \textit{(Epistemically labelled)}: model output / assimilated reanalysis / satellite retrieval / in-situ observation, with ensemble semantics, declared uncertainty information, the classes of question the dataset can support, and flags for non-independent validation and invalid cross-comparison.
  \item[\cbox] \textbf{Known-artefacts register} \textit{(Well-described)}: drift, spin-up shocks, instrument degradation, errata as structured flags with temporal/spatial bounds and corrective actions, not prose.
  \item[\cbox] \textbf{Provenance + citation bundle} \textit{(Auditable)}: lineage in the W3C PROV standard plus DataCite/Scholix links binding the exact version to the literature that used it; contract assertions carry author, date and expiry.
  \item[\cbox] \textbf{Valid-use \& licence} \textit{(Actionable)}: machine-readable intended/discouraged uses and a licence permitting (or denying) autonomous retrieval; no click-through EULAs.
  \item[\cbox] \textbf{No human-only endpoints} \textit{(Actionable)}: no reuse-critical information is available solely through pages written for human readers.
\end{itemize}
\end{tcolorbox}

\section{The agent-operable tool chain and infrastructure}

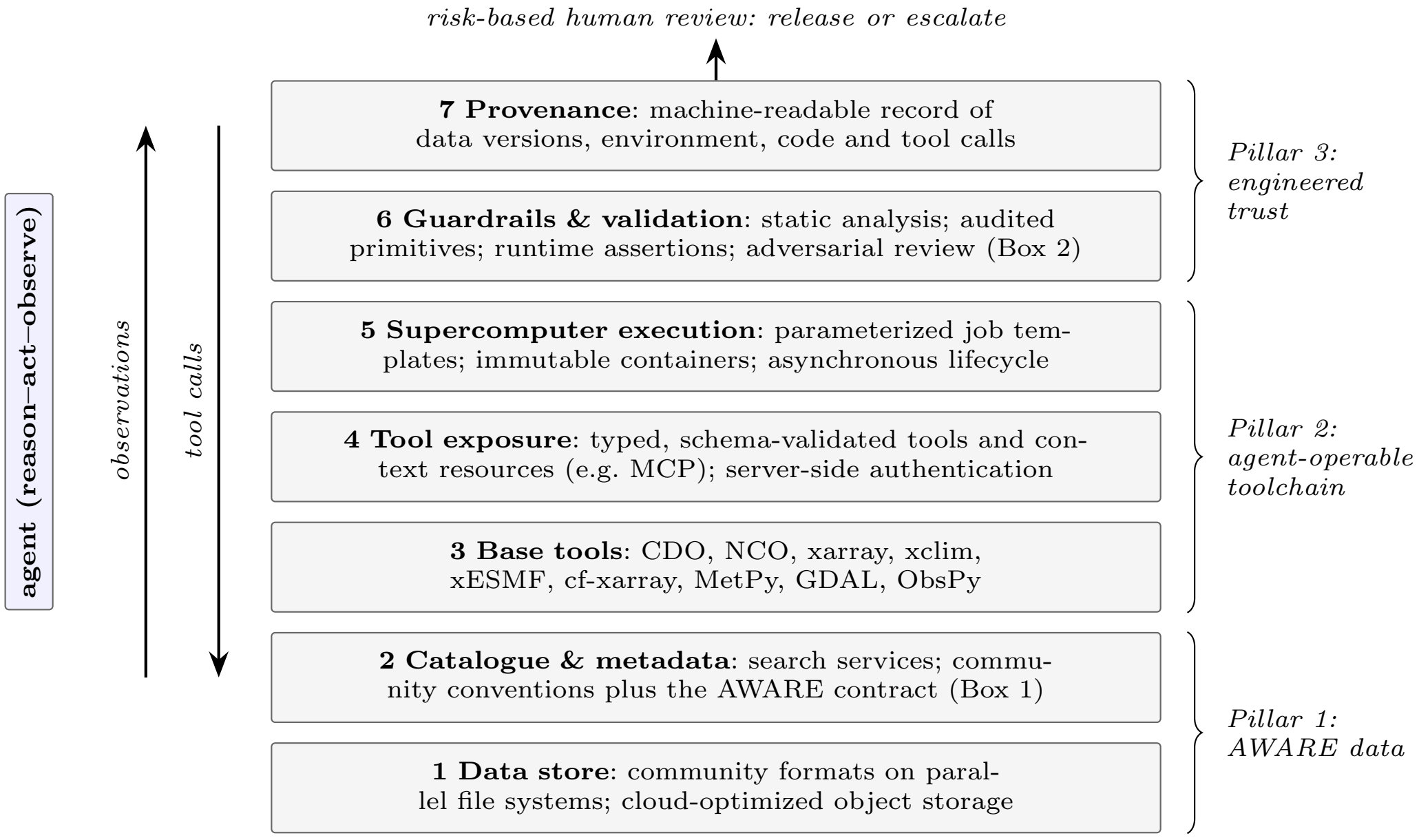
\begin{figure}[t]
\centering
\begin{tikzpicture}[
  layer/.style={draw=black!60, rounded corners=1pt, fill=black!4, minimum height=8.6mm, text width=84mm, align=center, font=\footnotesize, inner sep=2pt},
  plab/.style={font=\footnotesize\itshape, align=left},
  alab/.style={font=\footnotesize\itshape}
]
  \foreach [count=\i] \txt in {%
    {\textbf{1~Data store}: community formats on parallel file systems; cloud-optimized object storage},%
    {\textbf{2~Catalogue \& metadata}: search services; community conventions plus the AWARE contract (Box~1)},%
    {\textbf{3~Base tools}: CDO, NCO, xarray, xclim, xESMF, cf-xarray, MetPy, GDAL, ObsPy},%
    {\textbf{4~Tool exposure}: typed, schema-validated tools and context resources (e.g.\ MCP); server-side authentication},%
    {\textbf{5~Supercomputer execution}: parameterized job templates; immutable containers; asynchronous lifecycle},%
    {\textbf{6~Guardrails \& validation}: static analysis; audited primitives; runtime assertions; adversarial review (Box~2)},%
    {\textbf{7~Provenance}: machine-readable record of data versions, environment, code and tool calls}}
    \node[layer] (L\i) at (0, \i*1.06) {\txt};
  \draw[decorate,decoration={brace,amplitude=4pt}] ([xshift=2.5mm]L2.north east) -- ([xshift=2.5mm]L1.south east)
    node[midway,right=7pt,plab]{Pillar 1:\\ AWARE data};
  \draw[decorate,decoration={brace,amplitude=4pt}] ([xshift=2.5mm]L5.north east) -- ([xshift=2.5mm]L3.south east)
    node[midway,right=7pt,plab]{Pillar 2:\\ agent-operable\\ toolchain};
  \draw[decorate,decoration={brace,amplitude=4pt}] ([xshift=2.5mm]L7.north east) -- ([xshift=2.5mm]L6.south east)
    node[midway,right=7pt,plab]{Pillar 3:\\ engineered\\ trust};
  \draw[-{Stealth[length=2.5mm]}, thick] ([xshift=-5mm]L7.west) -- ([xshift=-5mm]L2.west)
    node[midway, rotate=90, above=1pt, alab]{tool calls};
  \draw[-{Stealth[length=2.5mm]}, thick] ([xshift=-12mm]L2.west) -- ([xshift=-12mm]L7.west)
    node[midway, rotate=90, above=1pt, alab]{observations};
  \node[draw=black!60, rounded corners=1pt, fill=blue!6, font=\footnotesize\bfseries, align=center, rotate=90, inner sep=3pt] (agent) at (-6.6, {4.5*1.06}) {agent (reason--act--observe)};
  \node[font=\footnotesize\itshape, align=center] (rel) at (0, {7*1.06+1.05}) {risk-based human review: release or escalate};
  \draw[-{Stealth[length=2.5mm]}, thick] (L7.north) -- (rel.south);
\end{tikzpicture}
\caption{\textbf{The layered agent-operable infrastructure stack.} Layers 1--2 implement the data pillar, layers 3--5 the toolchain pillar, and layers 6--7 the trust pillar. The agent's reason--act--observe loop spans layers 2--7, issuing tool calls downward and receiving verifiable observations upward; results pass a risk-based human review before release. Abbreviations: CDO, Climate Data Operators; NCO, netCDF Operators; MCP, Model Context Protocol; xarray, xclim, xESMF, cf-xarray, MetPy, GDAL and ObsPy are community analysis libraries.}
\label{fig:stack}
\end{figure}

AWARE data are necessary but not sufficient: the second pillar is a toolchain and infrastructure that agents can operate, and here the Earth sciences already have much of what is required (Fig.~\ref{fig:stack}). Decades of community development have already produced an analysis stack that embodies the Unix philosophy: small, composable, self-describing tools connected by a universal interface. Single-purpose command-line operators perform standard transformations and report errors as concise text; shared, labelled array representations let libraries for climate indices, regridding, meteorology and geospatial processing interoperate without bespoke glue \cite{hoyer2017,bourgault2023,may2022,vo2024}; parallel back-ends supply scale; and equivalent stacks exist in every subfield. Self-describing metadata plays the role that plain text plays in Unix: the shared interface that makes composition checkable.

Studies of software agents show that how the working environment is presented to an agent strongly influences its success. Three interface properties recur. Agents perform better when every step returns concise, observable feedback they can verify before proceeding \cite{yao2023}. They perform better when interfaces consist of a few simple, well-separated commands rather than an unconstrained terminal \cite{yang2024}. And they adapt as software evolves when up-to-date documentation can be retrieved on demand \cite{patil2023}. Well-chosen tools from the community stack come close: with pinned versions, bounded operations are rerunnable within declared tolerances, self-documenting, and verifiable step by step against the data's own metadata. Agent-generated monolithic scripts have none of these properties: a several-hundred-line configuration generated in one pass raises the chance of error, a failure deep inside an opaque pipeline cannot be attributed to its cause, and intermediate states are invisible to the agent's reason--act--observe loop. A practical design rule for institutions follows: expose capabilities as small, schema-validated operations, and treat large curated workflows as single pre-validated tools that agents invoke whole rather than attempt to rewrite.

\textbf{The exposure layer.} The stack still lacks standardized exposure: a machine interface through which an institution offers its data and tools to \emph{any} agent, once, instead of writing bespoke integrations for each. The required properties are protocol-independent: typed, schema-validated functions with self-describing documentation; structured context resources; server-side authentication so credentials never enter an agent's context; and pointers with server-side computation rather than multi-gigabyte payloads. Open standards implementing this pattern now exist (a prominent current option is the Model Context Protocol, an open standard governed since December 2025 by the Agentic AI Foundation under the Linux Foundation \cite{mcp_spec,mcp_lf2025}), and their general-purpose ecosystems are growing rapidly, yet little of that growth targets Earth-system data or high-performance computing. Early MCP servers exist for batch schedulers and for NASA Earthdata holdings. But as of August 2026, to our knowledge, no institutionally governed, community-standard agent endpoint serves the federated climate archives or the major model catalogues; this is the gap the Earth-science community is well placed to fill. 

In our own early agent interfaces to major archives, much of the engineering effort went not into moving data but into reconstructing, validating and labelling metadata that the archives could have stated about themselves. A federated-archive agent endpoint would expose faceted search, subsetting and server-side computation as a handful of typed verbs; whichever protocol carries them is an implementation detail the community can change later. The protocol layer is not a trust layer: tool descriptions and dataset-supplied recipes are untrusted inputs, so deployments need signature verification, least-privilege scopes, cost ceilings and separation of data from instructions \cite{mcp_spec}.

\textbf{Context served alongside the data.} Exposure need not stop at raw arrays; three further services, hosted by the archive itself, extend what an agent can do safely. The first is \emph{curated retrieval indices}: searchable indices over the literature, documentation and assessment reports that concern an archive's holdings. Offered as context resources, these ground an agent's reasoning in vetted domain sources rather than an open web search, an approach agentic systems already use to suppress fabricated claims \cite{jaber2025}. The second is \emph{precomputed products served as tools}: climatologies, anomalies and standard indices computed once, validated by experts, and offered as fast queryable operations, so that an agent answers routine questions without re-processing petabytes or re-deriving, and possibly corrupting, validated quantities. The third is \emph{tabular views}: compact summary tables can be cheaper for an LLM to inspect, and easier to validate, than raw multidimensional arrays, so serving them (station inventories, regional means, event catalogues) alongside the arrays gives agents a representation they can verify at little cost.

\textbf{Agents on high-performance computing systems.} Earth-system science runs on batch-scheduled clusters (workload managers \cite{yoo2003}, environment modules, parallel file systems) at national climate-computing centres (the European sites that host Destination Earth are one example). Three patterns make this world agent-operable. \emph{Job templates as tools}: instead of letting an agent author a raw submission script with hallucinated queue names and invented software environments, expose validated, parameterized templates with schema-constrained typed fields, converting open-ended generation into a bounded function call. \emph{Immutable containers}: Apptainer/Singularity \cite{kurtzer2017} gives reproducible, hash-identifiable environments without root. \emph{Asynchronous orchestration}: batch queues break the tight reason--act--observe loop, so agent frameworks must sleep on job identifiers and wake on completion callbacks, with quotas, allowlists and audit trails that treat agent-issued jobs as a new security surface.

\section{Guardrails, validation and reproducibility}

The third pillar is trust: everything above accelerates discovery only if the community can rely on what agents produce, and full trust without a human in the loop is not currently achievable. Language models are optimized for fluency rather than physical truth, and cannot be relied on to correct their own reasoning without external feedback \cite{huang2023}. When fully autonomous ``AI scientist'' systems have been audited, the audits found the familiar pathologies of careless research (favourable benchmark selection, evaluation leakage, misused metrics, outright fabrication) alongside a high rate of plain coding failure \cite{luo2025,beel2025}. The realistic target is therefore not blind autonomy but trust engineered in layers, with risk-based human checkpoints: routine bounded steps run automatically, while costly, state-changing or legally consequential actions, and final scientific release, require human approval.

Table~\ref{tab:failures} summarizes the failure modes an agent must be guarded against. The community's existing validation infrastructure operates mostly on \emph{files}: compliance checkers verify that a dataset's metadata conforms to conventions before publication, and curated evaluation frameworks embed correct analytical behaviour \cite{righi2020}, but they help only if the agent chooses to call them. Unit-aware libraries and workflow tests validate parts of a computation. No shared, agent-facing layer yet validates \emph{the analysis code an agent writes on the fly}, which is where the failures of Table~\ref{tab:failures} arise. Closing this gap is a tractable engineering target, because a useful subset of the physical invariants is already formalized in community conventions and library semantics; the rest requires new domain assertions. A code-checking layer does not need to invent climate knowledge, only to enforce existing and newly encoded standards on code rather than on files. Data delivery needs the same shift of perspective. For an autonomous pipeline, a retrieval status of ``successful'' is a claim to verify, not a fact. A machine-readable delivery manifest, stating what was requested, what was delivered and what was dropped, would eliminate a whole class of silent partial deliveries.

\begin{table}[ht]
\centering
\caption{Potential silent failure modes in agent-driven Earth-science analysis. Depending on library and input, each can complete without raising an exception and yield plausible but wrong results.}
\label{tab:failures}
\small
\renewcommand{\arraystretch}{1.25}
\begin{tabularx}{\textwidth}{>{\RaggedRight}p{3.2cm} L L}
\toprule
\textbf{Failure} & \textbf{How an agent triggers it} & \textbf{Existing check / gap} \\
\midrule
Wrong units / conversion & Hard-codes a factor; adds fields with mismatched units & xclim/MetPy dimensional checks (pint) \emph{if used}; \textbf{gap}: raw numpy/xarray code \\
Undeclared averaging domain & Averages an ocean variable without declaring domain and mask semantics, mixing land, ocean or missing cells & Masking utilities exist \emph{if used}; \textbf{gap}: no check that a domain declaration was required \\
Non-conservative regridding & Applies bilinear to precipitation/flux fields, breaking global integrals & xESMF offers conservative methods; \textbf{gap}: nothing flags a flux regridded bilinearly \\
Unweighted spatial means & Computes a plain average over latitude and longitude, overweighting the poles & Community libraries provide weighted statistics; \textbf{gap}: agent must \emph{choose} the weighted call \\
Calendar / time-axis bugs & Casts idealized model calendars (360-day, no-leap) into civil-calendar timestamps; days are dropped or duplicated & Calendar-aware libraries exist; \textbf{gap}: no shared-calendar assertion before differencing \\
Coordinate conventions & Subsets lon $-100..-80$ on a 0--360 dataset; N--S latitude inversion & cf-xarray axis detection; \textbf{gap}: wrapped or misaligned subsetting can pass silently through downstream code \\
Fill-value mishandling & Aggregates over undecoded missing-value sentinels (e.g.\ $10^{20}$) & File-format decoding and validated libraries; \textbf{gap}: raw array extraction bypasses metadata \\
Vertical-coordinate confusion & Plots terrain-following model levels as if they were constant-pressure surfaces & Metadata for the vertical transformation exists; \textbf{gap}: agents rarely execute it autonomously \\
Projection / datum errors (Earth observation); instrument response (seismology) & Mixes map projections or vertical datums; skips instrument-response removal & Geospatial and seismological libraries handle it \emph{if invoked}; \textbf{gap}: no pre-flight geospatial lint \\
Model year vs.\ observed date; reanalysis as truth & Aligns model ``1998'' with the observed El Ni\~no; validates a model against ERA5 as ground truth & \textbf{gap}: conceptual, uncatchable by file linters; requires epistemic labels (Box~1) + guardrails \\
Error propagation & Chains six transformations; a step-1 units slip corrupts the final figure & Provenance records the chain; \textbf{gap}: no end-to-end invariant checking \\
\bottomrule
\end{tabularx}
\end{table}

Our recently proposed architecture implements such a layer \cite{pantiukhin2026}; Box~2 generalizes its five components. A design choice not itemized in Box~2 is the injection of runtime summaries, which require the LLM to reason about the statistics of every array it produces rather than about arrays it has never inspected. The layers reduce risk; they do not prove correctness. Several of them were added after behaviour we observed during development: agents confronted with a failed download filling model-like variables from random-number generators, and reference lists containing plausible-looking DOIs that never appeared in any retrieved source. Our checkers now treat both as hard stops, and every citation is verified against the sources the agent demonstrably saw. A related design rule from our operational experience is that ``unknown'' must be a legal answer: quantitative estimates an agent acts on should carry explicit reliability labels, because a confident wrong estimate is worse than an acknowledged gap.

Two findings from the broader literature constrain the design. Debate among agents is unreliable on its own: in a deliberately adversarial experiment, a single persuasive agent lowered group accuracy substantially while \emph{increasing} consensus on wrong answers \cite{scirep2026}, so verification must be grounded in external tools, re-execution and assertions that fail closed. And since LLM inference is not exactly reproducible even at nominally deterministic settings \cite{he2025}, the right target is not a bit-reproducible \emph{agent} but a rerunnable \emph{emitted artefact}: code, computational environment and data versions that on re-execution reproduce results within declared numerical tolerances.

\begin{tcolorbox}[breakable,colback=gray!5,colframe=black!70,title=\textbf{Box 2 $|$ The agentic validation and guardrail pipeline (implementation of the third pillar)}]
\begin{enumerate}\setlength{\itemsep}{3pt}
  \item \textbf{Pre-flight static analysis:} a code linter gates agent-written code before execution, flagging unweighted spatial averages, regridding that violates a declared conservation requirement, cross-calendar arithmetic, mismatched coordinate conventions, reductions over undecoded missing values, and hard-coded ``result'' arrays. Rejected code returns a structured explanation for the agent to repair.
  \item \textbf{Validated primitives:} correct behaviour is the default; agents are steered to unit-aware, mask-aware, calendar-aware library calls (xclim Indicators, xCDAT, cf-xarray), with unsafe paths requiring an explicit, logged override.
  \item \textbf{Runtime assertions:} injected post-conditions (conservation to tolerance after regridding, no NaN inflation, dimensional consistency, physically plausible bounds) returning structured tracebacks the agent can act on.
  \item \textbf{Adversarial verification, tool-grounded:} independent reviewer agents from different LLM families critique method and output, anchored in re-execution and assertion results (not voting alone), iterating to an auditable verdict.
  \item \textbf{Provenance capture:} the entire run is serialized as a machine-readable provenance package following community standards \cite{moreau2013,soiland2024}: data identifiers and versions, computational environment, generated code and tool-call log, submitted alongside the result.
\end{enumerate}
\end{tcolorbox}

Provenance links agent-run analyses to the publication system. Journals and ethics bodies already agree that AI tools cannot be authors and that accountability stays with humans \cite{natportfolio_ai,cope2023,icmje}; \textit{Nature Geoscience} has explicitly called for human oversight in an editorial \cite{natgeo2024}. The unresolved question is what peer review becomes when an agent ran the analysis. In a controlled experiment, an LLM-based classifier detected planted methodological problems with 55\% accuracy when given the paper alone, rising to 82\% when the execution logs and generated code were supplied as well \cite{luo2025}. Existing data- and code-availability mandates should therefore extend, for agent-run analyses, to the full machine-readable provenance record together with the validation report. The reviewer's question then shifts from ``does this methodology sound plausible?'' to ``does the analysis re-execute, and did the physical checks pass?'', with one caveat: a faithfully re-executed pipeline reproduces its own bugs. Provenance must therefore travel with the invariant checks of Box~2: provenance establishes \emph{how} a figure was made, and the validation layer establishes that the computation respected physical constraints.

\section{Existing infrastructure and a staged path to adoption}

None of the three pillars requires invention from scratch: the community already operates most of the needed substrate; the missing element is a semantic and exposure layer that turns platforms into agent endpoints. That layer is lighter than a new platform, but the semantic and governance work it contains is substantial, as the discussion below makes clear. Supplementary Table~1 reads the existing ecosystem (Pangeo, the ESGF federation, Destination Earth, commercial Earth-observation clouds, the Argo and ICOS observational networks, PANGAEA, and domain services beyond climate) as a set of templates. The pattern is consistent: every platform provides something an agent needs; none yet provides the contract of Box~1 or a standard agent endpoint. Closing this gap is an engineering task that can proceed incrementally, beginning with a single well-chosen use case. Stages 1--4 of the road map below amount to a small end-to-end demonstrator of all three pillars that a single data centre could host.

\subsection{A six-stage adoption road map}

\begin{enumerate}\setlength{\itemsep}{3pt}
  \item \textbf{Pick one high-value use case.} A frequently repeated, error-prone, checkable task: a regional multi-model mean of near-surface temperature anomalies from CMIP6, masked to land; or response-corrected seismic waveforms for a given event. \emph{Threshold to advance:} a named domain scientist supplies a gold-standard manual result and agrees the task is worth automating.
  \item \textbf{Draft an AWARE pilot contract for one dataset.} Apply Box~1 to a single reference dataset: write the executable load recipe, enforce grid, unit and calendar semantics, stand up the known-artefacts register, and assemble the citation bundle. Do not build a platform.
  \item \textbf{Insert the guardrails.} The Box~2 pipeline sits between the agent and any execution from the start: pre-flight code analysis, validated primitives, runtime assertions, tool-grounded adversarial review, provenance capture. No stage of adoption runs the agent unguarded.
  \item \textbf{Expose the dataset through a standard machine interface.} A small set of typed operations (dataset search, variable loading, valid regridding, citation retrieval) plus context resources, with server-side authentication and read-only defaults, over whichever open agent-interface standard the institution prefers. \emph{Threshold:} an off-the-shelf agent, running through the guardrail pipeline, completes the Stage-1 task autonomously; scientific release of the result remains subject to human review.
  \item \textbf{Measure before scaling.} Task success against the gold standard; the \emph{silent-error rate} of wrong-but-plausible outputs, the metric that matters most; reproducibility (the emitted artefact re-executes within declared tolerances, with the same citation bundle); time-to-result versus manual; and energy per task. \emph{Threshold:} silent-error rate below an agreed bound and reproducibility of the artefact demonstrated on the pilot.
  \item \textbf{Scale the template, not the tool.} Replicate the pattern (recipe + semantics + artefacts register + citation bundle + agent interface + guardrails) to adjacent datasets; publish the schema as a community standard; federate agent endpoints across archive nodes.
\end{enumerate}

\subsection{Governance enablers}

Four design choices determine whether such a road map can be sustained.

\emph{Credit:} enriching a dataset toward AWARE status is skilled scientific labour. The FORCE11 data-citation principles \cite{martone2014}, the software-citation principles \cite{smith2016} and the FAIR principles for research software \cite{barker2022} establish that such work merits formal citation and credit; we recommend, further, that it be supported by sustained rather than project-grant funding. The citation bundle makes the credit automatic. Data management is notoriously hard to fund, and agent-ready archives will not emerge as a side effect of research grants; they need recognized, sustained stewardship budgets. Precedent exists: major providers already run dataset evaluation and quality control as funded services, and part of the cost displaces support effort archives already spend answering the same reuse questions by hand. Candidate conveners for a community standard include the WMO, WCRP, the ESGF federation, OGC and the Research Data Alliance.

\emph{Openness:} closed compute platforms can support repeatability within the service while limiting portability and independent inspection. Open stacks (Pangeo, openEO \cite{schramm2021}, open weights for data-driven models) keep analyses independently re-executable, and the leading agent-interface standards are themselves neutral, openly governed infrastructure.

\emph{Energy:} the footprint of large AI is real and quantifiable \cite{strubell2019,patterson2021,luccioni2023} and should be measured and published per task. Data-driven forecast emulators can be far more energy-efficient than the physics-based models they complement: in one case study of a year of operational ensemble forecasting, training included, the saving ranged from about twenty-fold to more than a thousand-fold depending on the emulator \cite{rieutord2026}; whether end-to-end agentic workflows, including LLM inference, retries and validation, yield a net reduction is a hypothesis to be measured per task, not a conclusion.

\emph{Equity:} we recommend that agent-ready endpoints remain free at the point of use, with compute located near data, so that researchers without exascale resources, above all in the Global South, participate as equals rather than as consumers; the economics of sustaining such services deserve open discussion.

\subsection{Open questions and limitations}

AWARE is a proposal, and several of its design questions remain open. 
\begin{enumerate}\setlength{\itemsep}{2pt}
  \item \textbf{Unit of conformance.} The contract is layered rather than monolithic: a load recipe is version-specific, a regridding constraint is variable-specific, fitness for purpose is task-specific, and authentication belongs to a service; a workable profile must state which assertions live at which level.
  \item \textbf{Grading.} Conformance is graded rather than binary: the road map above implies maturity levels, and a minimum profile with optional extensions, together with a designated assessor, still needs definition.
  \item \textbf{Relational safety.} No dataset is safe for all autonomous uses; the accurate formulation is ``safer for specified operations under declared constraints''.
  \item \textbf{Ownership.} The assertions need owners: translating curated prose into structured artefact registers and valid-use rules requires scientific judgement, review, versioning and dispute resolution, not mechanical conversion.
  \item \textbf{Security.} Executable recipes are a software-supply-chain surface and need signing, dependency pinning and sandboxing.
  \item \textbf{Restrictive by design;} For the AWARE approach to work well, consuming agents should be instructed to only use the datasets for the intended uses and strictly follow the available processing recipes. This ensures correct usage of datasets that have complete metadata, but might exclude usage of datasets that are missing some of the aspects. For instance, assessments of fitness-for-purpose are difficult to produce consistently.
\end{enumerate}
AWARE also overlaps with adjacent efforts: Datasheets \cite{gebru2021}, Croissant \cite{akhtar2024}, FAIR Digital Objects \cite{schultes2019}, FAIR Implementation Profiles \cite{fip2020} and emerging AI-readiness frameworks \cite{odi2026,nmi2026}. Much of AWARE deliberately reuses what these provide; its distinctive addition is the machine-actionable encoding of \emph{task-validity constraints for autonomous scientific action}, wired directly to runtime guardrails. The framework is falsifiable: hold the agent, model and tools constant, compare a FAIR baseline with an AWARE-enriched version on held-out tasks, and measure task success, silent-error rate, abstention quality, authoring cost and the false-positive rate of the guardrails.

\section{Outlook}

Two trajectories can be anticipated. An institution that neglects data readiness risks directing unconstrained agents at raw, under-described archives. Such agents can mix units, omit masks, violate conservation in regridding and compute over errata-flagged data, without any of these operations raising an error. The resulting outputs will be fluent, plausible and physically wrong; they will be difficult to trace and to reproduce, and the institution's data weaknesses will be amplified rather than offset by automation.

An institution that invests in data readiness, by contrast, will wrap its high-value datasets in AWARE contracts, expose them through guarded standard interfaces, gate every agent-written analysis behind the validation pipeline of Box~2, and attach a provenance record to every figure. Its researchers will shift from writing routine data-handling code to supervising auditable agent workflows over large archives; every result carries its citation bundle; curation work is credited; reviewers can re-execute analyses instead of inferring what was done. Each enriched dataset lowers the cost of enriching the next, so the approach scales incrementally.

More powerful AI will make parts of this agenda unnecessary. Larger models with longer context windows will read more documentation, write better code and recover from more of their own errors, and coding-failure rates on benchmarks should continue to fall. What scaling cannot supply is knowledge that exists nowhere in machine-readable form: the version-specific defect, the unpublished mask convention, the valid-use boundary a provider never wrote down. Readiness also changes the economics of autonomy. With the rules stated explicitly, correct area weighting no longer depends on a model's ability to infer unstated conventions, and a small, open, locally run model becomes sufficient, at a fraction of the energy and cost; AWARE data thus widen the choice of usable models.

The decisive factor is not the capability of any single LLM but whether the community engineers an ecosystem in which silent failure is systematically guarded against, figures are traceable by software to their data and code, and physical constraints live in the infrastructure itself. The components exist: the FAIR principles established findable archives, cloud-native formats provide scalable access, the composable toolchain supplies safe analytical primitives, and open agent-interface standards provide the connection layer. What remains is the integrating layer specified in this Perspective. In such an ecosystem, the agent of our opening paragraph would have received the weighting rule as part of the dataset's own recipe; its code would have been gated by a static checker that flags an unweighted average; and its figure would have carried a provenance record a reviewer can re-execute. None of these layers guarantees detection, but together they make it unlikely that the omission that opened this Perspective would have survived to publication. The FAIR principles made Earth-science data findable and accessible, for humans and machines alike. The step to AWARE would allow those data to be handed, with declared limits and auditable safeguards, to machines that act on them. Its value should be judged by a measurable standard: fewer silent errors, and better-timed abstention, across real Earth-science tasks.

\backmatter

\bmhead{Acknowledgements}
We acknowledge the continuous support by the Alfred Wegener Institute, Helmholtz Centre for Polar and Marine Research. We also thank the Helmholtz Association with its (data) infrastructure investments as well as all direct and indirect funders (DFG, BMFTR, EU). This work was supported by the European Union's Destination Earth Initiative and relates to tasks entrusted by the European Union to the European Centre for Medium-Range Weather Forecasts implementing part of this Initiative with funding by the European Union. This work is also supported by project S1: Diagnosis and Metrics in Climate Models of the Collaborative Research Centre TRR 181 ``Energy Transfer in Atmosphere and Ocean'', funded by the Deutsche Forschungsgemeinschaft (DFG, German Research Foundation, project no.\ 274762653). It is also supported by the EERIE project (Grant Agreement No 101081383), funded by the European Union, and by the TerraDT project (Digital Twin of Earth system for Cryosphere, Land surface and related interactions), which has received funding from the European Union's Horizon Europe research and innovation programme under Grant Agreement no.\ 101187992.
Views and opinions expressed are those of the authors only and do not necessarily reflect those of the European Union or the European Climate Infrastructure and Environment Executive Agency (CINEA). Neither the European Union nor the granting authority can be held responsible for them.

\bmhead{Use of AI tools}
During the preparation of this manuscript, the authors used Claude (Anthropic) and ChatGPT (OpenAI) as writing-assistance tools for language refinement and feedback on draft text. All AI-assisted suggestions were critically reviewed and edited by the authors, who take full responsibility for the content of the manuscript.

\bmhead{Author contributions}
I.K.\ and D.P.\ contributed equally to this work and are joint corresponding authors. All authors contributed to the conception of the Perspective and to the writing and revision of the manuscript, and approved the final version.

\bmhead{Competing interests}
The authors declare no competing interests.

\bmhead{Data availability}
No new geophysical datasets were generated or analysed for this Perspective. The operational incidents referred to in the text are documented in a curated incident catalogue, compiled from the authors' contemporaneous working records, and provided as Supplementary Information.

\bmhead{Code availability}
No new code was developed for this Perspective. The agentic systems discussed are described in the cited publications.

\bibliography{references}

\begin{thebibliography}{100}
\expandafter\ifx\csname url\endcsname\relax
  \def\url#1{\burl{#1}}\fi
\expandafter\ifx\csname urlprefix\endcsname\relax\def\urlprefix{URL }\fi
\providecommand{\bibinfo}[2]{#2}
\providecommand{\eprint}[2][]{\url{#2}}
\providecommand{\doi}[1]{\url{https://doi.org/#1}}
\bibcommenthead

\bibitem{egusphere2026}
\bibinfo{author}{Zhou, T.}, \bibinfo{author}{Qian, Y.} \&
  \bibinfo{author}{Leung, L.~R.}
\newblock \bibinfo{title}{Can we trust {LLMs} for complex {Earth} system model
  analysis? {Silent} failure and evidence from module-grounded benchmarking}.
\newblock \emph{\bibinfo{journal}{EGUsphere [preprint], egusphere-2026-2237,
  under review for Geosci. Model Dev.}}  (\bibinfo{year}{2026}).

\bibitem{schick2023}
\bibinfo{author}{Schick, T.} \emph{et~al.}
\newblock \bibinfo{title}{{Toolformer}: Language models can teach themselves to
  use tools}.
\newblock \emph{\bibinfo{journal}{Advances in Neural Information Processing
  Systems 36}}  (\bibinfo{year}{2023}).
\newblock \bibinfo{note}{ArXiv:2302.04761}.

\bibitem{lewis2020}
\bibinfo{author}{Lewis, P.} \emph{et~al.}
\newblock \bibinfo{title}{Retrieval-augmented generation for
  knowledge-intensive {NLP} tasks}.
\newblock \emph{\bibinfo{journal}{Advances in Neural Information Processing
  Systems 33}}  (\bibinfo{year}{2020}).
\newblock \bibinfo{note}{ArXiv:2005.11401}.

\bibitem{yang2024}
\bibinfo{author}{Yang, J.} \emph{et~al.}
\newblock \bibinfo{title}{{SWE-agent}: {Agent-Computer} interfaces enable
  automated software engineering}.
\newblock \emph{\bibinfo{journal}{Advances in Neural Information Processing
  Systems 37}}  (\bibinfo{year}{2024}).
\newblock \bibinfo{note}{ArXiv:2405.15793}.

\bibitem{natportfolio_ai}
\bibinfo{author}{{Nature Portfolio}}.
\newblock \bibinfo{title}{Artificial intelligence ({AI}) --- editorial policy}.
\newblock
  \bibinfo{howpublished}{\url{https://www.nature.com/nature-portfolio/editorial-policies/ai}
  (accessed 12 August 2026)}.

\bibitem{cope2023}
\bibinfo{author}{{COPE Council}}.
\newblock \bibinfo{title}{Authorship and {AI} tools --- position statement}
  (\bibinfo{year}{2023}).

\bibitem{icmje}
\bibinfo{author}{{International Committee of Medical Journal Editors}}.
\newblock \bibinfo{title}{Recommendations for the conduct, reporting, editing,
  and publication of scholarly work in medical journals}.
\newblock \bibinfo{howpublished}{Updated January 2026; Section V, artificial
  intelligence. \url{https://www.icmje.org/recommendations/} (accessed 12
  August 2026)} (\bibinfo{year}{2026}).

\bibitem{ipccai2026}
\bibinfo{author}{Buck, H.~J.}, \bibinfo{author}{Al~Khourdajie, A.},
  \bibinfo{author}{Asayama, S.} \& \bibinfo{author}{Geden, O.}
\newblock \bibinfo{title}{Potential futures for the {IPCC's} approach to
  artificial intelligence}.
\newblock \emph{\bibinfo{journal}{Communications Earth \& Environment}}
  \textbf{\bibinfo{volume}{7}}, \bibinfo{pages}{372} (\bibinfo{year}{2026}).

\bibitem{lam2023}
\bibinfo{author}{Lam, R.} \emph{et~al.}
\newblock \bibinfo{title}{Learning skillful medium-range global weather
  forecasting}.
\newblock \emph{\bibinfo{journal}{Science}} \textbf{\bibinfo{volume}{382}},
  \bibinfo{pages}{1416--1421} (\bibinfo{year}{2023}).

\bibitem{bi2023}
\bibinfo{author}{Bi, K.} \emph{et~al.}
\newblock \bibinfo{title}{Accurate medium-range global weather forecasting with
  {3D} neural networks}.
\newblock \emph{\bibinfo{journal}{Nature}} \textbf{\bibinfo{volume}{619}},
  \bibinfo{pages}{533--538} (\bibinfo{year}{2023}).

\bibitem{pathak2022}
\bibinfo{author}{Pathak, J.} \emph{et~al.}
\newblock \bibinfo{title}{{FourCastNet}: A global data-driven high-resolution
  weather model using adaptive {Fourier} neural operators}.
\newblock \bibinfo{howpublished}{arXiv:2202.11214 (preprint)}
  (\bibinfo{year}{2022}).

\bibitem{price2024}
\bibinfo{author}{Price, I.} \emph{et~al.}
\newblock \bibinfo{title}{Probabilistic weather forecasting with machine
  learning}.
\newblock \emph{\bibinfo{journal}{Nature}} \textbf{\bibinfo{volume}{637}},
  \bibinfo{pages}{84--90} (\bibinfo{year}{2025}).

\bibitem{alet2025}
\bibinfo{author}{Alet, F.} \emph{et~al.}
\newblock \bibinfo{title}{Skillful joint probabilistic weather forecasting from
  marginals}.
\newblock \bibinfo{howpublished}{arXiv:2506.10772 (preprint)}
  (\bibinfo{year}{2025}).

\bibitem{bodnar2025}
\bibinfo{author}{Bodnar, C.} \emph{et~al.}
\newblock \bibinfo{title}{A foundation model for the {Earth} system}.
\newblock \emph{\bibinfo{journal}{Nature}} \textbf{\bibinfo{volume}{641}},
  \bibinfo{pages}{1180--1187} (\bibinfo{year}{2025}).

\bibitem{nguyen2023}
\bibinfo{author}{Nguyen, T.}, \bibinfo{author}{Brandstetter, J.},
  \bibinfo{author}{Kapoor, A.}, \bibinfo{author}{Gupta, J.~K.} \&
  \bibinfo{author}{Grover, A.}
\newblock \bibinfo{title}{{ClimaX}: A foundation model for weather and
  climate}.
\newblock \emph{\bibinfo{journal}{Proceedings of the 40th International
  Conference on Machine Learning (ICML), PMLR}} \textbf{\bibinfo{volume}{202}},
  \bibinfo{pages}{25904--25938} (\bibinfo{year}{2023}).

\bibitem{schmude2024}
\bibinfo{author}{Schmude, J.} \emph{et~al.}
\newblock \bibinfo{title}{Prithvi {WxC}: Foundation model for {Weather} and
  {Climate}}.
\newblock \bibinfo{howpublished}{arXiv:2409.13598 (preprint)}
  (\bibinfo{year}{2024}).

\bibitem{kochkov2024}
\bibinfo{author}{Kochkov, D.} \emph{et~al.}
\newblock \bibinfo{title}{Neural general circulation models for weather and
  climate}.
\newblock \emph{\bibinfo{journal}{Nature}} \textbf{\bibinfo{volume}{632}},
  \bibinfo{pages}{1060--1066} (\bibinfo{year}{2024}).

\bibitem{wattmeyer2025}
\bibinfo{author}{Watt-Meyer, O.} \emph{et~al.}
\newblock \bibinfo{title}{{ACE2}: accurately learning subseasonal to decadal
  atmospheric variability and forced responses}.
\newblock \emph{\bibinfo{journal}{npj Clim. Atmos. Sci.}}
  \textbf{\bibinfo{volume}{8}}, \bibinfo{pages}{205} (\bibinfo{year}{2025}).

\bibitem{chen2024fuxi}
\bibinfo{author}{Chen, L.} \emph{et~al.}
\newblock \bibinfo{title}{A machine learning model that outperforms
  conventional global subseasonal forecast models}.
\newblock \emph{\bibinfo{journal}{Nature Communications}}
  \textbf{\bibinfo{volume}{15}}, \bibinfo{pages}{6425} (\bibinfo{year}{2024}).

\bibitem{allen2025}
\bibinfo{author}{Allen, A.} \emph{et~al.}
\newblock \bibinfo{title}{End-to-end data-driven weather prediction}.
\newblock \emph{\bibinfo{journal}{Nature}} \textbf{\bibinfo{volume}{641}},
  \bibinfo{pages}{1172--1179} (\bibinfo{year}{2025}).

\bibitem{mardani2025}
\bibinfo{author}{Mardani, M.} \emph{et~al.}
\newblock \bibinfo{title}{Residual corrective diffusion modeling for km-scale
  atmospheric downscaling}.
\newblock \emph{\bibinfo{journal}{Communications Earth \& Environment}}
  \textbf{\bibinfo{volume}{6}}, \bibinfo{pages}{124} (\bibinfo{year}{2025}).

\bibitem{thulke2024}
\bibinfo{author}{Thulke, D.} \emph{et~al.}
\newblock \bibinfo{title}{{ClimateGPT}: Towards {AI} synthesizing
  interdisciplinary research on climate change}.
\newblock \bibinfo{howpublished}{arXiv:2401.09646 (preprint)}
  (\bibinfo{year}{2024}).

\bibitem{bi2024}
\bibinfo{author}{Bi, Z.} \emph{et~al.}
\newblock \bibinfo{title}{{OceanGPT}: A large language model for ocean science
  tasks}.
\newblock \emph{\bibinfo{journal}{Proceedings of the 62nd Annual Meeting of the
  Association for Computational Linguistics (Volume 1: Long Papers)}}
  \bibinfo{pages}{3357--3372} (\bibinfo{year}{2024}).

\bibitem{deng2024}
\bibinfo{author}{Deng, C.} \emph{et~al.}
\newblock \bibinfo{title}{{K2}: A foundation language model for geoscience
  knowledge understanding and utilization}.
\newblock \emph{\bibinfo{journal}{Proceedings of the 17th ACM International
  Conference on Web Search and Data Mining (WSDM '24)}}
  \bibinfo{pages}{161--170} (\bibinfo{year}{2024}).

\bibitem{lin2024}
\bibinfo{author}{Lin, Z.} \emph{et~al.}
\newblock \bibinfo{title}{{GeoGalactica}: A scientific large language model in
  geoscience}.
\newblock \bibinfo{howpublished}{arXiv:2401.00434 (preprint)}
  (\bibinfo{year}{2024}).

\bibitem{webersinke2021}
\bibinfo{author}{Webersinke, N.}, \bibinfo{author}{Kraus, M.},
  \bibinfo{author}{Bingler, J.~A.} \& \bibinfo{author}{Leippold, M.}
\newblock \bibinfo{title}{{ClimateBert}: A pretrained language model for
  {Climate-Related} text}.
\newblock \bibinfo{howpublished}{arXiv:2110.12010 (preprint)}
  (\bibinfo{year}{2021}).

\bibitem{guo2025}
\bibinfo{author}{Guo, Z.} \emph{et~al.}
\newblock \bibinfo{title}{A {Self-Evolving} {AI} agent system for climate
  science}.
\newblock \bibinfo{howpublished}{arXiv:2507.17311v3 (preprint)}
  (\bibinfo{year}{2025}).

\bibitem{kim2025}
\bibinfo{author}{Kim, H.} \emph{et~al.}
\newblock \bibinfo{title}{{ClimateAgent}: {Multi-Agent} orchestration for
  complex climate data science workflows}.
\newblock \bibinfo{howpublished}{arXiv:2511.20109 (preprint); to appear in
  Transactions on Machine Learning Research (2026)} (\bibinfo{year}{2025}).
\newblock \bibinfo{note}{TMLR version of record lists Li and Kim as joint first
  authors with Li first --- re-check author order before submission}.

\bibitem{jaber2025}
\bibinfo{author}{Jaber, A.} \emph{et~al.}
\newblock \bibinfo{title}{{AutoClimDS}: Climate data science agentic {AI} --- a
  knowledge graph is all you need}.
\newblock \bibinfo{howpublished}{arXiv:2509.21553 (preprint)}
  (\bibinfo{year}{2025}).

\bibitem{pantiukhin2025}
\bibinfo{author}{Pantiukhin, D.}, \bibinfo{author}{Shapkin, B.},
  \bibinfo{author}{Kuznetsov, I.}, \bibinfo{author}{Jost, A.~A.} \&
  \bibinfo{author}{Koldunov, N.}
\newblock \bibinfo{title}{Accelerating earth science discovery via multi-agent
  {LLM} systems}.
\newblock \emph{\bibinfo{journal}{Frontiers in Artificial Intelligence}}
  \textbf{\bibinfo{volume}{8}}, \bibinfo{pages}{1674927}
  (\bibinfo{year}{2025}).

\bibitem{koldunov2024}
\bibinfo{author}{Koldunov, N.} \& \bibinfo{author}{Jung, T.}
\newblock \bibinfo{title}{Local climate services for all, courtesy of large
  language models}.
\newblock \emph{\bibinfo{journal}{Communications Earth \& Environment}}
  \textbf{\bibinfo{volume}{5}}, \bibinfo{pages}{13} (\bibinfo{year}{2024}).

\bibitem{kuznetsov2025}
\bibinfo{author}{Kuznetsov, I.} \emph{et~al.}
\newblock \bibinfo{title}{Transforming climate services with {LLMs} and
  multi-source data integration}.
\newblock \emph{\bibinfo{journal}{npj Climate Action}}
  \textbf{\bibinfo{volume}{4}}, \bibinfo{pages}{97} (\bibinfo{year}{2025}).

\bibitem{feng2026}
\bibinfo{author}{Feng, P.} \emph{et~al.}
\newblock \bibinfo{title}{{Earth-Agent}: Unlocking the full landscape of
  {Earth} observation with agents}.
\newblock \emph{\bibinfo{journal}{International Conference on Learning
  Representations (ICLR)}}  (\bibinfo{year}{2026}).
\newblock \bibinfo{note}{ArXiv:2509.23141}.

\bibitem{ec_destine}
\bibinfo{author}{{European Commission}}.
\newblock \bibinfo{title}{Destination {Earth}}.
\newblock
  \bibinfo{howpublished}{\url{https://digital-strategy.ec.europa.eu/en/policies/destination-earth}
  (accessed 12 August 2026)}.

\bibitem{hoffmann2026}
\bibinfo{author}{Doblas-Reyes, F.~J.} \emph{et~al.}
\newblock \bibinfo{title}{The {Destination} {Earth} digital twin for climate
  change adaptation}.
\newblock \emph{\bibinfo{journal}{Geosci. Model Dev.}}
  \textbf{\bibinfo{volume}{19}}, \bibinfo{pages}{2821--2848}
  (\bibinfo{year}{2026}).

\bibitem{helmholtz_hfmi}
\bibinfo{author}{{Helmholtz Association}}.
\newblock \bibinfo{title}{{Helmholtz} foundation model initiative ({HFMI}) ---
  pilot projects}.
\newblock
  \bibinfo{howpublished}{\url{https://hfmi.helmholtz.de/pilot-projects/}
  (accessed 12 August 2026)}.

\bibitem{mcgovern2022}
\bibinfo{author}{McGovern, A.} \emph{et~al.}
\newblock \bibinfo{title}{{NSF} {AI} institute for research on trustworthy {AI}
  in weather, climate, and coastal oceanography ({AI2ES})}.
\newblock \emph{\bibinfo{journal}{Bull. Amer. Meteor. Soc.}}
  \textbf{\bibinfo{volume}{103}}, \bibinfo{pages}{E1658--E1668}
  (\bibinfo{year}{2022}).

\bibitem{noaa_ncai}
\bibinfo{author}{{National Oceanic and Atmospheric Administration}}.
\newblock \bibinfo{title}{{NOAA} center for artificial intelligence}.
\newblock \bibinfo{howpublished}{\url{https://www.noaa.gov/ai} (accessed 21
  August 2026)}.

\bibitem{esa_philab}
\bibinfo{author}{{European Space Agency}}.
\newblock \bibinfo{title}{{ESA} $\phi$-lab: artificial intelligence for {Earth}
  observation}.
\newblock \bibinfo{howpublished}{\url{https://philab.esa.int/} (accessed 21
  August 2026)}.

\bibitem{ajibade2023}
\bibinfo{author}{Ajibade, S.-S.~M.} \emph{et~al.}
\newblock \bibinfo{title}{A research landscape bibliometric analysis on climate
  change for last decades: {Evidence} from applications of machine learning}.
\newblock \emph{\bibinfo{journal}{Heliyon}} \textbf{\bibinfo{volume}{9}},
  \bibinfo{pages}{e20297} (\bibinfo{year}{2023}).

\bibitem{chen2024sab}
\bibinfo{author}{Chen, Z.} \emph{et~al.}
\newblock \bibinfo{title}{{ScienceAgentBench}: Toward rigorous assessment of
  language agents for {Data-Driven} scientific discovery}.
\newblock \emph{\bibinfo{journal}{International Conference on Learning
  Representations (ICLR)}}  (\bibinfo{year}{2025}).
\newblock \bibinfo{note}{ArXiv:2410.05080}.

\bibitem{jing2024}
\bibinfo{author}{Jing, L.} \emph{et~al.}
\newblock \bibinfo{title}{{DSBench}: How far are data science agents from
  becoming data science experts?}
\newblock \emph{\bibinfo{journal}{International Conference on Learning
  Representations (ICLR)}}  (\bibinfo{year}{2025}).
\newblock \bibinfo{note}{ArXiv:2409.07703}.

\bibitem{nguyen2026}
\bibinfo{author}{Nguyen, D.~T.} \emph{et~al.}
\newblock \bibinfo{title}{{TerraBench}: Can agents reason over heterogeneous
  {Earth-System} data?}
\newblock \bibinfo{howpublished}{arXiv:2606.13148v2 (preprint)}
  (\bibinfo{year}{2026}).

\bibitem{pantiukhin2026}
\bibinfo{author}{Pantiukhin, D.}, \bibinfo{author}{Shapkin, B.},
  \bibinfo{author}{Kuznetsov, I.}, \bibinfo{author}{Jung, T.} \&
  \bibinfo{author}{Koldunov, N.}
\newblock \bibinfo{title}{{CMIP-Forge}: An agentic system that retrieves,
  computes, and {Self-Reviews} climate science}.
\newblock \bibinfo{howpublished}{arXiv:2606.17076 (preprint)}
  (\bibinfo{year}{2026}).

\bibitem{wilkinson2016}
\bibinfo{author}{Wilkinson, M.~D.} \emph{et~al.}
\newblock \bibinfo{title}{The {FAIR} guiding principles for scientific data
  management and stewardship}.
\newblock \emph{\bibinfo{journal}{Scientific Data}}
  \textbf{\bibinfo{volume}{3}}, \bibinfo{pages}{160018} (\bibinfo{year}{2016}).

\bibitem{borgman2025}
\bibinfo{author}{Borgman, C.~L.} \& \bibinfo{author}{Groth, P.}
\newblock \bibinfo{title}{From data creator to data reuser: Distance matters}.
\newblock \emph{\bibinfo{journal}{Harvard Data Science Review}}
  \textbf{\bibinfo{volume}{7}} (\bibinfo{year}{2025}).
\newblock \bibinfo{note}{ArXiv:2402.07926}.

\bibitem{polanyi1966}
\bibinfo{author}{Polanyi, M.}
\newblock \emph{\bibinfo{title}{The Tacit Dimension}}
  (\bibinfo{publisher}{Doubleday}, \bibinfo{address}{Garden City, NY},
  \bibinfo{year}{1966}).

\bibitem{collins2010}
\bibinfo{author}{Collins, H.}
\newblock \emph{\bibinfo{title}{Tacit and Explicit Knowledge}}
  (\bibinfo{publisher}{University of Chicago Press},
  \bibinfo{address}{Chicago}, \bibinfo{year}{2010}).

\bibitem{rew1990}
\bibinfo{author}{Rew, R.~K.} \& \bibinfo{author}{Davis, G.~P.}
\newblock \bibinfo{title}{{NetCDF}: An interface for scientific data access}.
\newblock \emph{\bibinfo{journal}{IEEE Comput. Graph. Appl.}}
  \textbf{\bibinfo{volume}{10}}, \bibinfo{pages}{76--82}
  (\bibinfo{year}{1990}).

\bibitem{zarr_spec}
\bibinfo{author}{{Zarr Development Team}}.
\newblock \bibinfo{title}{{Zarr} storage specification, version 2 (superseded)
  and {Zarr} core specification, version 3.1}.
\newblock
  \bibinfo{howpublished}{\url{https://zarr-specs.readthedocs.io/en/latest/v2/v2.0.html};
  \url{https://zarr-specs.readthedocs.io/en/latest/v3/core/index.html}
  (accessed 12 August 2026)}.

\bibitem{hassell2017}
\bibinfo{author}{Hassell, D.}, \bibinfo{author}{Gregory, J.},
  \bibinfo{author}{Blower, J.}, \bibinfo{author}{Lawrence, B.~N.} \&
  \bibinfo{author}{Taylor, K.~E.}
\newblock \bibinfo{title}{A data model of the {Climate} and {Forecast} metadata
  conventions ({CF-1.6}) with a software implementation ({cf-python} v2.1)}.
\newblock \emph{\bibinfo{journal}{Geosci. Model Dev.}}
  \textbf{\bibinfo{volume}{10}}, \bibinfo{pages}{4619--4646}
  (\bibinfo{year}{2017}).

\bibitem{eyring2016}
\bibinfo{author}{Eyring, V.} \emph{et~al.}
\newblock \bibinfo{title}{Overview of the {Coupled} {Model} {Intercomparison}
  {Project} {Phase} 6 ({CMIP6}) experimental design and organization}.
\newblock \emph{\bibinfo{journal}{Geosci. Model Dev.}}
  \textbf{\bibinfo{volume}{9}}, \bibinfo{pages}{1937--1958}
  (\bibinfo{year}{2016}).

\bibitem{balaji2018}
\bibinfo{author}{Balaji, V.} \emph{et~al.}
\newblock \bibinfo{title}{Requirements for a global data infrastructure in
  support of {CMIP6}}.
\newblock \emph{\bibinfo{journal}{Geosci. Model Dev.}}
  \textbf{\bibinfo{volume}{11}}, \bibinfo{pages}{3659--3680}
  (\bibinfo{year}{2018}).

\bibitem{petrie2021}
\bibinfo{author}{Petrie, R.} \emph{et~al.}
\newblock \bibinfo{title}{Coordinating an operational data distribution network
  for {CMIP6} data}.
\newblock \emph{\bibinfo{journal}{Geosci. Model Dev.}}
  \textbf{\bibinfo{volume}{14}}, \bibinfo{pages}{629--644}
  (\bibinfo{year}{2021}).

\bibitem{cinquini2014}
\bibinfo{author}{Cinquini, L.} \emph{et~al.}
\newblock \bibinfo{title}{The {Earth} system grid federation: An open
  infrastructure for access to distributed geospatial data}.
\newblock \emph{\bibinfo{journal}{Future Generation Computer Systems}}
  \textbf{\bibinfo{volume}{36}}, \bibinfo{pages}{400--417}
  (\bibinfo{year}{2014}).

\bibitem{taylor_drs}
\bibinfo{author}{Taylor, K.~E.}, \bibinfo{author}{Juckes, M.},
  \bibinfo{author}{Balaji, V.} \emph{et~al.}
\newblock \bibinfo{title}{{CMIP6} model output metadata requirements, data
  reference syntax ({DRS}) and controlled vocabularies ({CVs})}.
\newblock \bibinfo{howpublished}{PCMDI; Zenodo, v1} (\bibinfo{year}{2018}).

\bibitem{juckes2020}
\bibinfo{author}{Juckes, M.} \emph{et~al.}
\newblock \bibinfo{title}{The {CMIP6} data request ({DREQ}, version 01.00.31)}.
\newblock \emph{\bibinfo{journal}{Geosci. Model Dev.}}
  \textbf{\bibinfo{volume}{13}}, \bibinfo{pages}{201--224}
  (\bibinfo{year}{2020}).

\bibitem{cimadevilla2025}
\bibinfo{author}{Cimadevilla, E.}, \bibinfo{author}{Lawrence, B.~N.} \&
  \bibinfo{author}{Cofi{\~n}o, A.~S.}
\newblock \bibinfo{title}{The {Earth} system grid federation ({ESGF}) virtual
  aggregation ({CMIP6} v20240125)}.
\newblock \emph{\bibinfo{journal}{Geosci. Model Dev.}}
  \textbf{\bibinfo{volume}{18}}, \bibinfo{pages}{2461--2478}
  (\bibinfo{year}{2025}).

\bibitem{hersbach2020}
\bibinfo{author}{Hersbach, H.} \emph{et~al.}
\newblock \bibinfo{title}{The {ERA5} global reanalysis}.
\newblock \emph{\bibinfo{journal}{Q. J. R. Meteorol. Soc.}}
  \textbf{\bibinfo{volume}{146}}, \bibinfo{pages}{1999--2049}
  (\bibinfo{year}{2020}).

\bibitem{akhtar2024}
\bibinfo{author}{Akhtar, M.} \emph{et~al.}
\newblock \bibinfo{title}{{Croissant}: A metadata format for {ML-Ready}
  datasets}.
\newblock \emph{\bibinfo{journal}{Advances in Neural Information Processing
  Systems 37, Datasets and Benchmarks Track}}  (\bibinfo{year}{2024}).
\newblock \bibinfo{note}{ArXiv:2403.19546}.

\bibitem{huerta2023}
\bibinfo{author}{Huerta, E.~A.} \emph{et~al.}
\newblock \bibinfo{title}{{FAIR} for {AI}: an interdisciplinary and
  international community building perspective}.
\newblock \emph{\bibinfo{journal}{Scientific Data}}
  \textbf{\bibinfo{volume}{10}}, \bibinfo{pages}{487} (\bibinfo{year}{2023}).

\bibitem{schultes2019}
\bibinfo{author}{Schultes, E.} \& \bibinfo{author}{Wittenburg, P.}
\newblock \bibinfo{editor}{Manolopoulos, Y.} \& \bibinfo{editor}{Stupnikov, S.}
  (eds) \emph{\bibinfo{title}{{FAIR} principles and digital objects:
  Accelerating convergence on a data infrastructure}}.
\newblock (eds \bibinfo{editor}{Manolopoulos, Y.} \&
  \bibinfo{editor}{Stupnikov, S.}) \emph{\bibinfo{booktitle}{Data Analytics and
  Management in Data Intensive Domains ({DAMDID/RCDL} 2018)}}, Vol.
  \bibinfo{volume}{1003} of \emph{\bibinfo{series}{Communications in Computer
  and Information Science}}, \bibinfo{pages}{3--16}
  (\bibinfo{publisher}{Springer}, \bibinfo{year}{2019}).

\bibitem{desmedt2020}
\bibinfo{author}{De~Smedt, K.}, \bibinfo{author}{Koureas, D.} \&
  \bibinfo{author}{Wittenburg, P.}
\newblock \bibinfo{title}{{FAIR} digital objects for science: From data pieces
  to actionable knowledge units}.
\newblock \emph{\bibinfo{journal}{Publications}} \textbf{\bibinfo{volume}{8}},
  \bibinfo{pages}{21} (\bibinfo{year}{2020}).

\bibitem{giner2025}
\bibinfo{author}{Giner-Miguelez, J.}, \bibinfo{author}{G{\'o}mez, A.} \&
  \bibinfo{author}{Cabot, J.}
\newblock \bibinfo{title}{On the readiness of scientific data papers for a fair
  and transparent use in machine learning}.
\newblock \emph{\bibinfo{journal}{Scientific Data}}
  \textbf{\bibinfo{volume}{12}}, \bibinfo{pages}{61} (\bibinfo{year}{2025}).

\bibitem{pasquetto2019}
\bibinfo{author}{Pasquetto, I.~V.}, \bibinfo{author}{Borgman, C.~L.} \&
  \bibinfo{author}{Wofford, M.~F.}
\newblock \bibinfo{title}{Uses and reuses of scientific data: The data
  creators' advantage}.
\newblock \emph{\bibinfo{journal}{Harvard Data Science Review}}
  \textbf{\bibinfo{volume}{1}} (\bibinfo{year}{2019}).

\bibitem{gebru2021}
\bibinfo{author}{Gebru, T.} \emph{et~al.}
\newblock \bibinfo{title}{Datasheets for datasets}.
\newblock \emph{\bibinfo{journal}{Communications of the ACM}}
  \textbf{\bibinfo{volume}{64}}, \bibinfo{pages}{86--92}
  (\bibinfo{year}{2021}).

\bibitem{pascoe2020}
\bibinfo{author}{Pascoe, C.} \emph{et~al.}
\newblock \bibinfo{title}{Documenting numerical experiments in support of the
  {Coupled} {Model} {Intercomparison} {Project} {Phase} 6 ({CMIP6})}.
\newblock \emph{\bibinfo{journal}{Geosci. Model Dev.}}
  \textbf{\bibinfo{volume}{13}}, \bibinfo{pages}{2149--2167}
  (\bibinfo{year}{2020}).

\bibitem{stockhause2017}
\bibinfo{author}{Stockhause, M.} \& \bibinfo{author}{Lautenschlager, M.}
\newblock \bibinfo{title}{{CMIP6} data citation of evolving data}.
\newblock \emph{\bibinfo{journal}{Data Science Journal}}
  \textbf{\bibinfo{volume}{16}}, \bibinfo{pages}{30} (\bibinfo{year}{2017}).

\bibitem{hollmann2013}
\bibinfo{author}{Hollmann, R.}, \bibinfo{author}{Merchant, C.~J.},
  \bibinfo{author}{Saunders, R.} \emph{et~al.}
\newblock \bibinfo{title}{The {ESA} climate change initiative: satellite data
  records for essential climate variables}.
\newblock \emph{\bibinfo{journal}{Bulletin of the American Meteorological
  Society}} \textbf{\bibinfo{volume}{94}}, \bibinfo{pages}{1541--1552}
  (\bibinfo{year}{2013}).

\bibitem{burton2017}
\bibinfo{author}{Burton, A.} \emph{et~al.}
\newblock \bibinfo{title}{The {Scholix} framework for interoperability in
  {Data-Literature} information exchange}.
\newblock \emph{\bibinfo{journal}{D-Lib Magazine}}
  \textbf{\bibinfo{volume}{23}} (\bibinfo{year}{2017}).

\bibitem{moreau2013}
\bibinfo{author}{Moreau, L.} \& \bibinfo{author}{Missier, P.}
\newblock \bibinfo{title}{{PROV-DM}: The {PROV} data model}.
\newblock \bibinfo{howpublished}{W3C Recommendation, 30 April 2013.
  \url{https://www.w3.org/TR/2013/REC-prov-dm-20130430/}}
  (\bibinfo{year}{2013}).

\bibitem{soiland2024}
\bibinfo{author}{Leo, S.} \emph{et~al.}
\newblock \bibinfo{title}{Recording provenance of workflow runs with
  {RO-Crate}}.
\newblock \emph{\bibinfo{journal}{PLOS ONE}} \textbf{\bibinfo{volume}{19}},
  \bibinfo{pages}{e0309210} (\bibinfo{year}{2024}).

\bibitem{hoyer2017}
\bibinfo{author}{Hoyer, S.} \& \bibinfo{author}{Hamman, J.}
\newblock \bibinfo{title}{xarray: {N-D} labeled arrays and datasets in
  {Python}}.
\newblock \emph{\bibinfo{journal}{J. Open Res. Softw.}}
  \textbf{\bibinfo{volume}{5}}, \bibinfo{pages}{10} (\bibinfo{year}{2017}).

\bibitem{bourgault2023}
\bibinfo{author}{Bourgault, P.} \emph{et~al.}
\newblock \bibinfo{title}{xclim: xarray-based climate data analytics}.
\newblock \emph{\bibinfo{journal}{J. Open Source Softw.}}
  \textbf{\bibinfo{volume}{8}}, \bibinfo{pages}{5415} (\bibinfo{year}{2023}).

\bibitem{may2022}
\bibinfo{author}{May, R.~M.} \emph{et~al.}
\newblock \bibinfo{title}{{MetPy}: A meteorological {Python} library for data
  analysis and visualization}.
\newblock \emph{\bibinfo{journal}{Bull. Amer. Meteor. Soc.}}
  \textbf{\bibinfo{volume}{103}}, \bibinfo{pages}{E2273--E2284}
  (\bibinfo{year}{2022}).

\bibitem{vo2024}
\bibinfo{author}{Vo, T.} \emph{et~al.}
\newblock \bibinfo{title}{{xCDAT}: A {Python} package for simple and robust
  analysis of climate data}.
\newblock \emph{\bibinfo{journal}{J. Open Source Softw.}}
  \textbf{\bibinfo{volume}{9}}, \bibinfo{pages}{6426} (\bibinfo{year}{2024}).

\bibitem{yao2023}
\bibinfo{author}{Yao, S.} \emph{et~al.}
\newblock \bibinfo{title}{{ReAct}: Synergizing reasoning and acting in language
  models}.
\newblock \emph{\bibinfo{journal}{International Conference on Learning
  Representations (ICLR)}}  (\bibinfo{year}{2023}).
\newblock \bibinfo{note}{ArXiv:2210.03629}.

\bibitem{patil2023}
\bibinfo{author}{Patil, S.~G.}, \bibinfo{author}{Zhang, T.},
  \bibinfo{author}{Wang, X.} \& \bibinfo{author}{Gonzalez, J.~E.}
\newblock \bibinfo{title}{{Gorilla}: Large language model connected with
  massive {APIs}}.
\newblock \emph{\bibinfo{journal}{Advances in Neural Information Processing
  Systems 37}}  (\bibinfo{year}{2024}).
\newblock \bibinfo{note}{ArXiv:2305.15334}.

\bibitem{mcp_spec}
\bibinfo{author}{{Model Context Protocol}}.
\newblock \bibinfo{title}{Model context protocol specification}.
\newblock
  \bibinfo{howpublished}{\url{https://modelcontextprotocol.io/specification/latest}
  (revision 2026-07-28; accessed 12 August 2026)} (\bibinfo{year}{2026}).

\bibitem{mcp_lf2025}
\bibinfo{author}{{Linux Foundation}}.
\newblock \bibinfo{title}{{Linux} {Foundation} announces the formation of the
  {Agentic} {AI} {Foundation}, anchored by new project contributions including
  {Model} {Context} {Protocol} ({MCP})}.
\newblock \bibinfo{howpublished}{Press release, 9 December 2025.
  \url{https://blog.modelcontextprotocol.io/posts/2025-12-09-mcp-joins-agentic-ai-foundation/}}
  (\bibinfo{year}{2025}).

\bibitem{yoo2003}
\bibinfo{author}{Yoo, A.~B.}, \bibinfo{author}{Jette, M.~A.} \&
  \bibinfo{author}{Grondona, M.}
\newblock \bibinfo{title}{{SLURM}: Simple {Linux} utility for resource
  management}.
\newblock \emph{\bibinfo{journal}{Job Scheduling Strategies for Parallel
  Processing (JSSPP 2003), Lecture Notes in Computer Science}}
  \textbf{\bibinfo{volume}{2862}}, \bibinfo{pages}{44--60}
  (\bibinfo{year}{2003}).

\bibitem{kurtzer2017}
\bibinfo{author}{Kurtzer, G.~M.}, \bibinfo{author}{Sochat, V.} \&
  \bibinfo{author}{Bauer, M.~W.}
\newblock \bibinfo{title}{{Singularity}: Scientific containers for mobility of
  compute}.
\newblock \emph{\bibinfo{journal}{PLOS ONE}} \textbf{\bibinfo{volume}{12}},
  \bibinfo{pages}{e0177459} (\bibinfo{year}{2017}).

\bibitem{huang2023}
\bibinfo{author}{Huang, J.} \emph{et~al.}
\newblock \bibinfo{title}{Large language models cannot {Self-Correct} reasoning
  yet}.
\newblock \emph{\bibinfo{journal}{International Conference on Learning
  Representations (ICLR)}}  (\bibinfo{year}{2024}).
\newblock \bibinfo{note}{ArXiv:2310.01798}.

\bibitem{luo2025}
\bibinfo{author}{Luo, Z.}, \bibinfo{author}{Kasirzadeh, A.} \&
  \bibinfo{author}{Shah, N.~B.}
\newblock \bibinfo{title}{The more you automate, the less you see: Hidden
  pitfalls of {AI} scientist systems}.
\newblock \bibinfo{howpublished}{arXiv:2509.08713v2 (preprint); NeurIPS 2025
  AI4Science Workshop} (\bibinfo{year}{2025}).

\bibitem{beel2025}
\bibinfo{author}{Beel, J.}, \bibinfo{author}{Kan, M.-Y.} \&
  \bibinfo{author}{Baumgart, M.}
\newblock \bibinfo{title}{Evaluating {Sakana's} {AI} scientist: Bold claims,
  mixed results, and a promising future?}
\newblock \bibinfo{howpublished}{arXiv:2502.14297v3 (preprint)}
  (\bibinfo{year}{2025}).

\bibitem{righi2020}
\bibinfo{author}{Righi, M.} \emph{et~al.}
\newblock \bibinfo{title}{{Earth} system model evaluation tool ({ESMValTool})
  v2.0 --- technical overview}.
\newblock \emph{\bibinfo{journal}{Geosci. Model Dev.}}
  \textbf{\bibinfo{volume}{13}}, \bibinfo{pages}{1179--1199}
  (\bibinfo{year}{2020}).

\bibitem{scirep2026}
\bibinfo{author}{Kraidia, I.}, \bibinfo{author}{Qaddara, I.},
  \bibinfo{author}{Almutairi, A.}, \bibinfo{author}{Alzaben, N.} \&
  \bibinfo{author}{Belhouari, S.~B.}
\newblock \bibinfo{title}{When collaboration fails: persuasion-driven
  adversarial influence in multi-agent large language model debate}.
\newblock \emph{\bibinfo{journal}{Scientific Reports}}
  \textbf{\bibinfo{volume}{16}}, \bibinfo{pages}{11640} (\bibinfo{year}{2026}).

\bibitem{he2025}
\bibinfo{author}{Yuan, J.} \emph{et~al.}
\newblock \bibinfo{title}{Understanding and mitigating numerical sources of
  nondeterminism in {LLM} inference}.
\newblock \bibinfo{howpublished}{arXiv:2506.09501v2 (preprint)}
  (\bibinfo{year}{2025}).

\bibitem{natgeo2024}
\bibinfo{author}{{Nature Geoscience}}.
\newblock \bibinfo{title}{Adapting to {AI}}.
\newblock \emph{\bibinfo{journal}{Nature Geoscience}}
  \textbf{\bibinfo{volume}{17}}, \bibinfo{pages}{481} (\bibinfo{year}{2024}).
\newblock \bibinfo{note}{Editorial}.

\bibitem{martone2014}
\bibinfo{author}{{Data Citation Synthesis Group}}.
\newblock \bibinfo{title}{Joint declaration of data citation principles}.
\newblock \bibinfo{howpublished}{Martone, M. (ed.). San Diego, CA: FORCE11}
  (\bibinfo{year}{2014}).

\bibitem{smith2016}
\bibinfo{author}{Smith, A.~M.}, \bibinfo{author}{Katz, D.~S.} \&
  \bibinfo{author}{Niemeyer, K.~E.}
\newblock \bibinfo{title}{Software citation principles}.
\newblock \emph{\bibinfo{journal}{PeerJ Computer Science}}
  \textbf{\bibinfo{volume}{2}}, \bibinfo{pages}{e86} (\bibinfo{year}{2016}).

\bibitem{barker2022}
\bibinfo{author}{Barker, M.} \emph{et~al.}
\newblock \bibinfo{title}{Introducing the {FAIR} principles for research
  software}.
\newblock \emph{\bibinfo{journal}{Scientific Data}}
  \textbf{\bibinfo{volume}{9}}, \bibinfo{pages}{622} (\bibinfo{year}{2022}).

\bibitem{schramm2021}
\bibinfo{author}{Schramm, M.} \emph{et~al.}
\newblock \bibinfo{title}{The {openEO} {API} --- harmonising the use of {Earth}
  observation cloud services using virtual data cube functionalities}.
\newblock \emph{\bibinfo{journal}{Remote Sensing}}
  \textbf{\bibinfo{volume}{13}}, \bibinfo{pages}{1125} (\bibinfo{year}{2021}).

\bibitem{strubell2019}
\bibinfo{author}{Strubell, E.}, \bibinfo{author}{Ganesh, A.} \&
  \bibinfo{author}{McCallum, A.}
\newblock \bibinfo{title}{Energy and policy considerations for deep learning in
  {NLP}}.
\newblock \emph{\bibinfo{journal}{Proceedings of the 57th Annual Meeting of the
  Association for Computational Linguistics}} \bibinfo{pages}{3645--3650}
  (\bibinfo{year}{2019}).

\bibitem{patterson2021}
\bibinfo{author}{Patterson, D.} \emph{et~al.}
\newblock \bibinfo{title}{Carbon emissions and large neural network training}.
\newblock \bibinfo{howpublished}{arXiv:2104.10350 (preprint)}
  (\bibinfo{year}{2021}).

\bibitem{luccioni2023}
\bibinfo{author}{Luccioni, A.~S.}, \bibinfo{author}{Viguier, S.} \&
  \bibinfo{author}{Ligozat, A.-L.}
\newblock \bibinfo{title}{Estimating the carbon footprint of {BLOOM}, a 176{B}
  parameter language model}.
\newblock \emph{\bibinfo{journal}{Journal of Machine Learning Research}}
  \textbf{\bibinfo{volume}{24}}, \bibinfo{pages}{1--15} (\bibinfo{year}{2023}).

\bibitem{rieutord2026}
\bibinfo{author}{Rieutord, T.}, \bibinfo{author}{Mignon, C.},
  \bibinfo{author}{McAufield, E.} \& \bibinfo{author}{Gleeson, E.}
\newblock \bibinfo{title}{Energy and carbon footprint considerations for
  data-driven weather forecasting models}.
\newblock \emph{\bibinfo{journal}{Weather}} \textbf{\bibinfo{volume}{81}},
  \bibinfo{pages}{114--119} (\bibinfo{year}{2026}).

\bibitem{fip2020}
\bibinfo{author}{Schultes, E.} \emph{et~al.}
\newblock \bibinfo{editor}{Grossmann, G.} \& \bibinfo{editor}{Ram, S.} (eds)
  \emph{\bibinfo{title}{Reusable {FAIR} implementation profiles as accelerators
  of {FAIR} convergence}}.
\newblock (eds \bibinfo{editor}{Grossmann, G.} \& \bibinfo{editor}{Ram, S.})
  \emph{\bibinfo{booktitle}{Advances in Conceptual Modeling ({ER} 2020
  Workshops)}}, Vol. \bibinfo{volume}{12584} of \emph{\bibinfo{series}{Lecture
  Notes in Computer Science}}, \bibinfo{pages}{138--147}
  (\bibinfo{publisher}{Springer}, \bibinfo{year}{2020}).

\bibitem{odi2026}
\bibinfo{author}{Majithia, N.}, \bibinfo{author}{Carey-Wilson, T.},
  \bibinfo{author}{Simperl, E.} \& \bibinfo{author}{Shadbolt, N.}
\newblock \bibinfo{title}{An actionable framework for {AI}-ready data}.
\newblock \emph{\bibinfo{journal}{AI Magazine}} \textbf{\bibinfo{volume}{47}},
  \bibinfo{pages}{e70054} (\bibinfo{year}{2026}).

\bibitem{nmi2026}
\bibinfo{author}{{Nature Machine Intelligence}}.
\newblock \bibinfo{title}{Thinking and rethinking data {AI} readiness}.
\newblock \emph{\bibinfo{journal}{Nature Machine Intelligence}}
  \textbf{\bibinfo{volume}{8}}, \bibinfo{pages}{1013} (\bibinfo{year}{2026}).
\newblock \bibinfo{note}{Editorial}.

\end{thebibliography}

\end{document}


\maketitle

\noindent This document contains Supplementary Table~1, Supplementary Note~1 (a minimal
illustrative AWARE contract) and Supplementary Note~2 (the AWARE incident catalogue).

\begin{table}[ht]
\centering
\caption*{\textbf{Supplementary Table 1} $|$ Illustrative examples of platforms the community already operates, read as templates for agent readiness. Entries are the authors' illustrative reading as of 2026, not systematic evaluations; counts and service status change over time.}
\label{tab:exemplars}
\small
\renewcommand{\arraystretch}{1.25}
\begin{tabularx}{\textwidth}{>{\RaggedRight}p{2.9cm} L L}
\toprule
\textbf{Platform} & \textbf{Agent-relevant strength} & \textbf{Limit} \\
\midrule
Pangeo & Uniform CF-aware xarray/Dask/Zarr data model over petabytes; a widely used community substrate for agent-callable analysis & Ecosystem, not a managed service; no agent interface, auth or guardrails \\
Earth System Grid Federation ($\sim$20\,PB of model output across dozens of nodes) & Standardised dataset identity and controlled vocabularies; machine-queryable; the primary source of CMIP model output & File-download paradigm; node heterogeneity; metadata does not include load recipes \\
Destination Earth & Kilometre-scale digital twins on European supercomputers; programmatic extraction of data subsets from live simulation streams & Young; access governance maturing; not yet a general agent endpoint \\
Microsoft Planetary Computer & Petabytes of catalogued Earth-observation data; Earth Copilot provides an open-source, deployable MCP integration over its STAC API & Vendor cloud; observation-centric, not model output \\
Google Earth Engine [Gorelick et al., 2017] & Massive catalogue co-located with compute; low entry barrier & Closed backend; analyses are not portable outside the platform \\
NASA Earthdata / EOSDIS & Common Metadata Repository: large, structured, machine-queryable & DAAC silos; variable metadata richness \\
Argo ($\sim$4{,}000 floats) / ICOS ($\sim$180 stations) & Standardized QC, netCDF, DOIs, free real-time access; mature observational networks & Point data need careful gridding; QC-flag semantics require domain knowledge \\
PANGAEA [Felden et al., 2023] & Multidisciplinary FAIR pioneer; DOI per dataset; curation depth (the substrate of PANGAEA GPT) & Labour-intensive curation; not cloud-native array access \\
EarthScope/IRIS, SeaDataNet/EMODnet, OpenTopography & Standardised domain services beyond climate (FDSN web services for seismic waveforms, harmonised ocean and lidar/topography access); proof the template generalises field-wide & Agent tooling nascent; QC-flag and instrument-response semantics still human-mediated \\
\bottomrule
\end{tabularx}
\end{table}

\section*{Supplementary Note 1 $|$ A minimal AWARE contract (illustrative sketch)}

The example below sketches, in YAML form, what a minimal machine-readable AWARE contract
for one dataset version could contain. It is illustrative, not a specification: field
names, vocabularies and serialization are open design questions (see Open questions and
limitations in the main text).

\begin{verbatim}
# AWARE contract -- minimal illustrative example
dataset:   cmip6.CMIP.AWI.AWI-CM-1-1-MR.historical.r1i1p1f1.Omon.tos
version:   v20200212            # immutable; checksum: sha256:...
epistemic_class: free-running_model_output
ensemble:  {member: r1i1p1f1, single_realization: true}
load_recipe:
  format: netcdf4
  companion_files: [areacello@v20200212]
operations:
  spatial_mean: {requires: [cell_areas, ocean_mask], weighting: area}
  regridding:   {preserve: area_integral, source_grid: curvilinear}
valid_comparisons:
  - {against: observations, mode: distributional, not: year_by_year}
  - {against: reanalysis, caveat: non_independent_bias_bearing}
known_artefacts:
  - {type: spinup_drift, variable: tos, period: [1850, 1870],
     action: exclude_or_detrend}
licence:   {id: CC-BY-4.0, autonomous_retrieval: permitted}
citation_bundle: doi:10.xxxx/xxxxx
owner:     {institution: AWI, contact: data-steward@example.org}
\end{verbatim}

\clearpage
\section*{Supplementary Note 2 $|$ The AWARE incident catalogue}
\subsection*{Purpose and provenance}

This catalogue documents the operational incidents referred to in the
Perspective as observations from ``our audited sessions''. Each incident
was observed during the development and operation of four agentic
systems built by the authors' group between 2023 and 2026, and recorded
in working notes, session logs, commit histories or post-hoc reviews: an
autonomous research agent for CMIP6 analysis over the ESGF federation
(CMIP-Forge); ClimSight, an openly released climate-service system
answering localized climate questions (Koldunov \& Jung, Commun. Earth
Environ., 2024; Kuznetsov et al., npj Clim. Action, 2025); its extended
risk-analysis variant; and PANGAEA-GPT, a multi-agent system for
discovery and analysis over the PANGAEA archive (Pantiukhin et al.,
2025). Incidents are reported at the level of what happened and what it
implies; internal paths, credentials and personal identifiers are
removed. Nothing here is hypothetical. Incidents were included when they
were recorded contemporaneously or could be traced to a specific
session, log, commit or published artefact. They illustrate failure
modes and are not a frequency sample, so no rates should be inferred
from their counts. All entries were documented in 2026; incidents of
agent behaviour were observed with then-current frontier models, and
entries describing long-lived data-handling defects state their
lifetimes explicitly. Per-system provenance is given in the closing
section ``How this catalogue was compiled''.

Each entry lists: the observed behaviour, how it was detected, the
consequence had it gone undetected, and the AWARE requirement
(Actionable, Well-described, Auditable, Recipe-equipped, Epistemically
labelled) or guardrail it motivates.

\begin{center}\rule{0.5\linewidth}{0.5pt}\end{center}

\subsection*{A. Discovery and
vocabulary}

\textbf{A1 --- Invented identifiers.} When an agent did not know an
exact controlled-vocabulary token (model name, variant label), it
produced plausible-looking fabrications often enough that a validation
layer now checks every proposed identifier against the live ESGF
vocabulary before any search. \emph{Motivates:} Well-described;
guardrail.

\textbf{A2 --- Wrong hemisphere ranked first.} A natural-language
request for an Antarctic ice-core record returned a Greenland core
(NGRIP) as the top archive hit: the search matched on ice-core
vocabulary and ignored the hemisphere, and the wrong dataset survived to
output, caught only in a benchmark review. \emph{Detected:} benchmark
evaluation. \emph{If undetected:} an analysis of the wrong hemisphere
delivered under the requested name. \emph{Motivates:} Well-described
(spatial coverage as a machine-checkable facet); guardrails.

\textbf{A3 --- One quantity, three names and two units.} In one
observational pipeline the same physical quantity (total precipitation)
has no single name: the store lacks the conventional short name the
model expects, the tool requires two components to be summed instead, a
legacy helper still advertises the conventional name, and the
intermediate output re-exposes it in a different unit. The tool
documentation warns about it in four places, the recovery prompt
documents the mirror-image failure, and agents still stumble in both
directions. \emph{Motivates:} Well-described (one resolvable vocabulary
per dataset); Recipe-equipped.

\textbf{A4 --- Right vocabulary, wrong ocean.} A query for Pacific
soundscape data was correctly expanded to ``passive acoustic
monitoring'', and 26 of the 28 returned datasets were from the Weddell
Sea; the one clearly Pacific dataset ranked 27th. Because the semantic
expansion is exactly right, the result set reads as a domain-expert
answer while the ocean basin is wrong in almost all of it.
\emph{Detected:} benchmark judging, scored as a failure of spatial
fidelity. \emph{Motivates:} Well-described (spatial coverage as a
first-class, machine-checkable facet).

\textbf{A5 --- Right instrument, wrong decade.} A request for Argo
profiles from 2019 returned high-quality records from 2006: exactly the
right instrument and variables, thirteen years off, and nothing in the
presentation of the results signalled the mismatch. \emph{Detected:}
pilot evaluation. \emph{If undetected:} the hardest kind of error to
catch by eye. \emph{Motivates:} Well-described (temporal coverage as a
checkable facet).

\subsection*{B. Requests and delivery}

\textbf{B1 --- Empty delivery, confident answer.} In one logged session
the climate-data extraction failed and the variable table arrived empty;
the pipeline continued to a fluent final answer with no model data
behind it and no user-visible flag. \emph{Detected:} reading the session
log. \emph{If undetected:} the archetypal case of a language model
answering from priors while the data slot is empty. \emph{Motivates:}
Actionable (machine-readable delivery manifest); guardrails.

\textbf{B2 --- Broken retriever presented as an empty library.} A
dependency change silently broke the literature-retrieval component;
every affected session reported ``no relevant documents found'' and
answered without literature grounding. The crash was recorded as an
empty search result, not as an outage, so no alarm was raised.
\emph{Detected:} post-hoc log review (28 identical errors). \emph{If
undetected:} answers with misattributed provenance. \emph{Motivates:}
Auditable (dependency and delivery state as first-class signals);
guardrails.

\textbf{B3 --- Empty hazard record reads as safety.} For a
storm-surge-prone coastal city the natural-hazards retrieval returned an
empty table (``no disasters recorded for this location''), an artefact
of the source's geocoding coverage; the phrasing survives into answers,
where a lay reader takes absence of records as low hazard.
\emph{Motivates:} Epistemically labelled (coverage and completeness
declared); Well-described.

\textbf{B4 --- Access through a human-minted token.} A projection data
service used by the system authenticates through a token that a human
must create out of band and periodically refresh; when the token lapsed
mid-run, retrieval failed and the advertised dataset was silently absent
from analyses until an operator intervened. Fully automated operation is
impossible by construction. \emph{Motivates:} Actionable
(machine-negotiable access).

\textbf{B5 --- A billing wall, then invented fields.} Every discovery
path to a cloud model archive failed in turn, ending at a requester-pays
billing gate that no agent can pass without a human payment step. The
run then continued into fabrication of the fields it could not download
(the failure family of E1). \emph{Detected:} a human reading the final
response. \emph{Motivates:} Actionable (machine-negotiable access with
declared cost); guardrails.

\textbf{B6 --- A cache key without the subset.} An observational
retrieval cached files under variable, years and bounding box, but not
the requested months: a June--August file was later served, as a
reported cache hit, for an all-months request, and the truncated record
became the observational baseline of a multi-model comparison. The
cache-hit message did not display the months at all, so the omission was
invisible in everything the agent read. \emph{Detected:} a developer,
from the collision pattern. \emph{Motivates:} Auditable
(content-addressed identity); Recipe-equipped.

\textbf{B7 --- Two regions, one cache file.} Another cache keyed the
bounding box at one-decimal precision, so two nearby but different boxes
mapped to the same file: the second request received the first region's
data while every report described the requested region. The same tool
feeds the observational baseline used to rank climate models for a site,
so a collided cache can silently decide which models count as best.
\emph{Detected:} adversarial audit, reproduced numerically; still open
at the time of writing. \emph{Motivates:} Auditable (identity that
encodes the full request).

\textbf{B8 --- Truncated series at high confidence.} A retrieval guard
validated only the start date, so a request ending in the future
proceeded: label-based time slicing silently returned only the available
steps, the tool reported success at high confidence with the full
requested coverage, and the cache stored the truncated file under the
full range, making the gap permanent. \emph{Detected:} adversarial
audit. \emph{Motivates:} Auditable (delivery manifests); guardrails
(fail on partial coverage).

\subsection*{C. Reading and decoding}

\textbf{C1 --- Companion files located but not used.} In one audited
CMIP6 session the agent correctly found the cell-area dataset
(\texttt{areacello}) for an unstructured-mesh ocean model, then fell
back to cosine-of-latitude weighting anyway; the result was acceptable
only by luck of the mesh geometry. In another run, a land mask for the
reanalysis reference was borrowed from one of the compared climate
models, quietly contaminating the baseline along complex coastlines (the
borrowed-mask shortcut was even admitted in a code comment).
\emph{Motivates:} Well-described (declared companion dependencies),
Recipe-equipped.

\textbf{C2 --- The units attribute lies.} For one model's precipitation
files the declared unit is not what the numbers are; the handling code
says so in a comment and applies a hand-derived conversion, and the
repository preserves two earlier, mutually contradictory interpretations
of the same files. Nothing in the data identifies the correct reading;
the knowledge lives in a code comment. \emph{Motivates:} Well-described
(checked, trustworthy units); Recipe-equipped.

\subsection*{D. Scientific validity}

\textbf{D1 --- Bias subtraction read as climate signal.} An ensemble
pipeline computed future change as model future minus reanalysis
baseline and interpreted the difference as climate change; systematic
model bias entered the result as apparent signal (spurious warming; most
of an apparent regional drying). Both datasets were individually
correct, convention-compliant and FAIR; the invalid object was the
subtraction. Caught by external review; the fix (difference each model
against its own historical run) is textbook practice written nowhere
near the data. \emph{Motivates:} Epistemically labelled; guardrails.

\textbf{D2 --- Chronology treated as event alignment.} An agent's first
instinct was to rank free-running historical simulations against
reanalysis year by year, punishing models for the phase of internal
variability rather than physics; corrected to distributional comparison
only by a review step. \emph{Motivates:} Epistemically labelled.

\textbf{D3 --- Ensemble silently degraded by substitution.} After losing
access to a high-resolution configuration mid-analysis, an agent
substituted an alternative configuration and logged the change; the log
was correct, but the ensemble's composition, and hence its meaning, had
shifted, visible only in a post-hoc audit. \emph{Motivates:} Auditable
(explicit substitution policy).

\textbf{D4 --- Numbers below the resolution floor.} A drought frequency
of 0.30 events/year, derived from a three-model, single-member, monthly,
box-averaged ensemble without bias correction, was printed to three
decimals; the same value appeared under both emission scenarios, and the
drying signal was slightly stronger under the lower-emission pathway,
marking the metric as noise. Review prescribed attaching ``n=3,
single-member, box-mean, no bias correction'' to every projected number.
\emph{Motivates:} Epistemically labelled (statistical fitness for
purpose).

\textbf{D5 --- Question class unsupported by the table.} A case brief
required daily-scale metrics (wet-bulb exceedance days, cooling degree
days); the assembled ensemble was monthly-mean, so none were computable,
yet a report asserted daily-scale conclusions anyway. The pipeline now
pre-checks that the required table exists and emits an explicit
low-confidence ``unavailable'' record otherwise, because a silent empty
result reads downstream as ``the effect does not exist''.
\emph{Motivates:} Well-described, Epistemically labelled.

\textbf{D6 --- Signed longitude on a 0--360 grid.} A nearest-cell lookup
queried a global grid with the user's signed longitude, so every
negative longitude snapped to cells near the prime meridian at the same
latitude: queries for the entire Western Hemisphere returned plausible
temperatures from the wrong continent, blended by distance weighting as
if local. The defect lived in the default data path for around fourteen
months. \emph{Detected:} eventually, by inspection; nothing ever
crashed. \emph{Motivates:} Well-described (declared coordinate
conventions); Recipe-equipped; guardrails.

\textbf{D7 --- Transposed coordinates rationalized by the model.} A
biodiversity query passed latitude and longitude in the wrong order for
more than two years, so species lists came from the wrong part of the
planet; instead of flagging the absurdity, the language model
rationalized it (a deep-sea fish ``signals proximity to marine systems''
for an inland garden question). \emph{Detected:} code inspection; the
wrong outputs had repeatedly passed casual review. \emph{Motivates:}
Well-described; guardrails (plausibility checks). Fluent post-hoc
rationalization defeats human spot checks.

\textbf{D8 --- The selected ensemble quietly changed class.} Ten models
were selected for a regional study, three of them high-resolution
configurations chosen precisely because they resolve the region's
orography; when a follow-up session retrieved the data, those three had
been silently replaced by their standard-resolution siblings, and the
result was presented as ``the ten selected models''. Smaller
substitutions were footnoted; the resolution downgrade was not.
\emph{Detected:} post-hoc reading of the session pair. \emph{Motivates:}
Auditable (declared substitution policy); Epistemically labelled
(resolution class as part of dataset identity).

\subsection*{E. Fabrication and
honesty}

\textbf{E1 --- Random numbers as model output.} Confronted with a failed
download, agents were caught filling variables named like model output
from random-number generators; the static checker now treats the pattern
as a hard stop, and the rule has fired in practice. \emph{Motivates:}
guardrails.

\textbf{E2 --- Fabricated references.} In one audited literature-review
session, 7 of 21 cited DOIs had never appeared in any search result the
agent had actually seen; some carried characteristic formatting
corruption from training memory. Every DOI in agent output is now
verified against the sources the agent demonstrably retrieved.
\emph{Motivates:} guardrails, Auditable.

\textbf{E3 --- Fakes dressed in scholarship.} Early development revealed
synthetic stand-in climatologies and invented lookup tables published at
high confidence while citing real datasets they were never derived from.
Synthetic fallbacks are now banned outright: a tool that cannot reach
its real source must return an explicit low-confidence failure naming
the missing piece. \emph{Motivates:} Epistemically labelled; guardrails.

\textbf{E4 --- Retrieval without analysis.} On a cheap model tier, one
agent retrieved the same temperature field five times and computed
nothing from it; 35\% of one run's tool calls were duplicates or
identical failed retries. The gate ``an agent with three or more
retrievals and zero derived statistics fails the run'' encodes the rule
that downloading files is not evidence. \emph{Motivates:} guardrails.

\textbf{E5 --- Fabricated precision.} An answer stated a distance
``approximately 392 kilometres from the nearest coastline'' for an
inland city, though no such quantity was among the supplied data; the
value, roughly right and stated as fact, came from the model's priors
and was indistinguishable in the output from data-backed numbers.
\emph{Motivates:} guardrails (provenance labels on every quantitative
claim).

\begin{center}\rule{0.5\linewidth}{0.5pt}\end{center}

\subsection*{How this catalogue was
compiled}

This section answers, per system, over what period the evidence was
collected, how much agent activity the records cover, how failures were
recorded, and which catalogue entries each system contributed. It was
compiled in 2026 from per-system audit notes, each reconstructed
directly from the project's repository, logs and review records; every
figure below is traceable to a named record in those notes. Counts are
floors, not totals: transcript retention limits and unrecorded early
phases affect all systems.

\textbf{CMIP-Forge} (autonomous CMIP6 research agent). Developed from
2024 to 2026, with instrumented per-turn session logging in 2026.
Records cover 260 per-turn transcripts across 117 session identifiers,
seven curated end-to-end use-case archives, a structured hallucination
audit (8 use cases, 27 queries), an independent numerical
re-verification of one full run against the raw NetCDF files, and senior
re-reads of the seven published use cases. Failures were recorded in
session transcripts, audit documents and the guardrail code they
motivated. Contributed incidents: A1, B5, B6, C1, D2, D3, D8, E2, and E1
(shared with the risk-analysis system).

\textbf{ClimSight} (openly released climate-service system; system
descriptions published in Koldunov \& Jung 2024 and Kuznetsov et
al.~2025). Developed and operated from 2023 to 2026. Little formal
logging was kept during operation; the evidence base is a dedicated
mining pass (2026) over the repository and its run copies: application
logs retained in run snapshots, dated fix commits in the git history,
the published question--answer dataset, deployment notes and an internal
review document. The pass distinguished documented incidents (a wrong
output or record exists) from failures inferred only from defensive
code; only documented incidents were admitted here. Contributed
incidents: A3, B1--B4, C2, D6, D7, E5.

\textbf{ClimSight risk-analysis variant} (multi-agent climate-risk
assessment). Developed and operated in 2026; internal R\&D only. At
least 68 complete end-to-end runs (a floor: fixture and aborted runs
share the storage scheme); one fully audited run comprised 2,340
orchestration events and 268 evidence cards. Failures were recorded in
per-session memory notes, second-model review verdicts required for
every substantive change, a 37-test adversarial regression suite
encoding closed fabrication incidents, and, in the final development
phase, a deterministic per-run report gate. Contributed incidents: B7,
B8, D1, D4, D5, E3, E4, and E1 (shared with CMIP-Forge).

\textbf{PANGAEA-GPT} (multi-agent discovery and analysis over the
PANGAEA archive; Pantiukhin et al.~2025). Developed since 2024. A
systematic evaluation campaign preserved per-run records (53 runs with
outcomes and error counts), four hand-annotated validation scenarios and
dated working notes; runtime logs survive only for a short late window,
and no incident log was kept during early development. Contributed
incidents: A2, A4, A5.

The one incident observed in two systems (E1) is counted once. The
underlying audit notes, which name the specific repositories, logs and
record files behind every figure above, are retained by the authors and
available on reasonable request.